\documentclass[%
reprint,
twocolumn,
nofootinbib,
 amsmath,amssymb,
]{revtex4-1}
\usepackage{graphicx}   
\usepackage{subfig}
\expandafter\let\csname equation*\endcsname\relax
\expandafter\let\csname endequation*\endcsname\relax
\usepackage{mathrsfs,amsmath}
\usepackage{siunitx} 
\usepackage{natbib}
\usepackage{array}
\usepackage[left=19mm, right = 19mm, top =18.5 mm, bottom=18.5mm]{geometry}
\usepackage[justification=raggedright]{caption}

\graphicspath{{figures/}}

\usepackage[us]{datetime}
\usepackage{fancyhdr}

\usepackage{hyperref}
\hypersetup{
    bookmarks=false,         
    unicode=false,          
    pdftoolbar=true,        
    pdfmenubar=true,        
    pdffitwindow=false,     
    pdftitle={Topical Review: Recent advances in mechanical torque studies of small-scale magnetism},    
    pdfauthor={J. E. Losby, V. T. K. Sauer, M. R. Freeman},     
    pdfsubject={},   
    pdfcreator={},   
    pdfproducer={},  
    pdfkeywords={}, 
    pdfnewwindow=true,      
    colorlinks=true,       
    linkcolor=black,          
    citecolor=black,        
    filecolor=black,      
    urlcolor=black          
}

\newcommand{\FSR}{F\!S\!R}
\newcommand{\FWHM}{FW\!H\!M}
\newcommand{\lt}{<}
\newcommand{\gt}{>}

\makeatletter

\renewcommand\subsubsection{\@startsection{subsubsection}{3}{\z@}%
                                     {-3.25ex\@plus -1ex \@minus -.2ex}%
                                     {1em \@plus .2em}%
                                     {\reset@font\normalsize\itshape\raggedright}} 
\makeatother

\makeatletter
\newcommand \Dotfill {\leavevmode \cleaders \hb@xt@ .8em{\hss .\hss }\hfill \kern \z@}
\makeatother

\makeatletter
\def\p@subsection{}
\def\p@subsubsection{}
\makeatother

\fancypagestyle{customstyle}{%
  \fancyhead[L]{\emph{Recent Advances in Mechanical Torque Studies of Small-scale Magnetism (Topical Review)}}
  \fancyhead[R]{\thepage}
  \fancyfoot{}
 }

\begin{document}

\title[Topical Review: Recent Advances in Mechanical Torque Studies of Small-scale Magnetism]{Topical Review: Recent advances in mechanical torque studies of small-scale magnetism}

\author{Joseph E. Losby, Vincent T. K. Sauer \& Mark R. Freeman}

\address{Department of Physics, University of Alberta, Edmonton, Alberta T6G 2E1, Canada}

\email[Corresponding author: ]{mark.freeman@ualberta.ca}


\date{\today}

\begin{abstract}
There is a storied scientific history in the role of mechanical instruments for the measurement of fundamental physical interactions.  Among these include the detection of magnetic torques via a displacement of a compliant mechanical sensor as a result of angular momentum transfer.  Modern nanofabrication methods have enabled the coupling of mechanical structures to single, miniature magnetic specimens.  This has allowed for strikingly sensitive detection of magnetic hysteresis and other quasi-static effects, as well as spin resonances, in materials confined to nanoscale geometries.  The extraordinary sensitivities achieved in mechanical transduction through recent breakthroughs in cavity optomechanics, where a high-finesse optical cavity is used for readout of motion, are now being harnessed for torque magnetometry.   In this article, we review the recent progress in mechanical detection of magnetic torques, highlight current applications, and speculate on possible future developments in the technology and science.  Guidelines for designing and implementing the measurements are also included. 
\end{abstract}

\maketitle

\pagestyle{customstyle}

\tableofcontents

\vfill

\begin{figure*}[ht]
\addcontentsline{toc}{section}{\protect\numberline{}List of Symbols}
\centering
\caption*{\bf{LIST OF SYMBOLS}}
\begin{minipage}[t]{0.45\linewidth}
\begin{tabular}[t]{>{\raggedright\arraybackslash}p{1.75cm} >{\raggedright\arraybackslash}p{5.5cm} }

$T$ & Temperature\\

DC & Direct current (field)\\

AC & Alternating current (field)\\

RF & Radio-frequency field\\

$\kappa$ & Spring constant\\

$\kappa_{\mathrm{eff}}$ & Effective spring constant\\

$\kappa_{\mathrm{inner}}$ & Torsional spring constant for the inner paddles in a symmetric triple-paddle resonator\\

$\kappa_{\mathrm{outer}}$ & Torsional spring constant for the inner paddles in a symmetric triple-paddle resonator\\

$I$ & Moment of inertia\\

$I_{\mathrm{eff}}$ & Effective moment of inertia\\

$I_{\mathrm{inner}}$ & Moment of inertia for the inner paddles in a symmetric triple-paddle resonator\\

$I_{\mathrm{outer}}$ & Moment of inertia for the outer paddles in a symmetric triple-paddle resonator\\

$b$ & Dimension of a side of a rectangular cross section of a torsion rod\\

$c$ & Dimension of a side of a rectangular cross section of a torsion rod\\ 
 
$L$ &Length of torsion rod\\
 
$G$ & Shear modulus\\
 
$\beta$ & Geometric correction factor for non-uniform shear stress around the perimeter of a rectangular cross-section of a torsion rod\\
 
$\rho_{\mathrm{area}}$ & Areal density\\
 
$f$ & Modal frequency\\
 
$f_0$ & Mechanical resonance frequency\\
 
$Q$ & Mechanical quality factor\\
 
$k_{\mathrm{B}}$ & Boltzmann constant\\
  
$\hbar$ & Planck's constant\\
 
$\Omega$ & Mechanical angular frequency\\
 
$n_{\mathrm{eff}}$ & effective index of refraction of the optical mode\\
 
$m$ & Optical mode number\\

$\omega_{\mathrm{opt}}$ & Optical resonant frequency\\

$L_{\mathrm{opt}}$ & Round trip length of light in an optical cavity\\

$Q_{\mathrm{opt}}$ & Optical quality factor\\

$\FSR$& Free spectral range\\

\end{tabular}
\end{minipage}
\begin{minipage}[t]{0.45\linewidth}
\begin{tabular}[t]{>{\raggedright\arraybackslash}p{1.75cm} >{\raggedright\arraybackslash}p{5.5cm} }

$\kappa_{\mathrm{opt}}$ & \mbox{Optical cavity decay rate;} \mbox{$\kappa_{\mathrm{opt}}=\kappa_{\mathrm{ex}}+\kappa_{\mathrm{0}}$} \\

$\kappa_{\mathrm{ex}}$ & Optical cavity input loss rate\\

$\kappa_{\mathrm{0}}$ & Optical cavity internal loss rate\\

$\lambda$ & Free space wavelength\\

$n_{\mathrm{g}}$ & Group index\\

$\mathcal{F}$ & Optical cavity finesse\\

$\FWHM$&Full width half max\\

$E$ & Electric field strength of laser\\

$\Delta$ & Probe laser detuning compared to the cavity\\

$\omega_\mathrm{p}$ & Free-space laser probe frequency\\

$\omega_\mathrm{cav}$ & Optical cavity frequency\\

$\Omega_\mathrm{mech}$ & Mechanical device angular frequency\\

$T_{\mathrm{opt}}$ & Optical transmission\\

$\phi$ & Phase of optical cavity\\

$z$ & Optomechanical beam displacement\\

$\lambda_\mathrm{p}$ & Free space laser probe wavelength\\

$T_0$ & On-resonance cavity transmission value\\

$L_\mathrm{mech}$ & Mechanical beam optical cavity interaction length\\

$g_\mathrm{om}$ & Optomechanical coupling constant $\partial \omega_\mathrm{opt} / \partial z$\\

$\tau$ & Magnetic torque\\
 
$M$ & Magnetic moment\\

$M_\mathrm{s}$ & Saturation magnetization\\

$\mu_0$ & Permeability of free space\\

$H_{\mathrm{x}},H_{\mathrm{y}},H_{\mathrm{z}}$ & x, y, and z components of field\\

$g \prime$ & Magnetomechanical ratio\\

$A$ & Aspect ratio of pendulum bob torsional resonator, $h/r$\\

$r$ & Torsion rod radius\\

$h$ & Torsion bob height, and diameter/width\\

$S_{\mathrm{\theta}}$ & Angular amplitude spectral density\\

$S_{\mathrm{z}}$ & Displacement amplitude spectral density\\

$R$ & Reflectance\\

EBL & Electron beam lithography\\

SOI & Silicon-on-insulator\\

FIB & Focussed ion beam\\

YIG & Yttrium iron garnet\\

SNR & Signal-to-noise ratio\\

\end{tabular}
\end{minipage}
\end{figure*}

\section{Introduction}

\begin{figure*}[!ht] 
	\centering
	\includegraphics[width=1\linewidth] {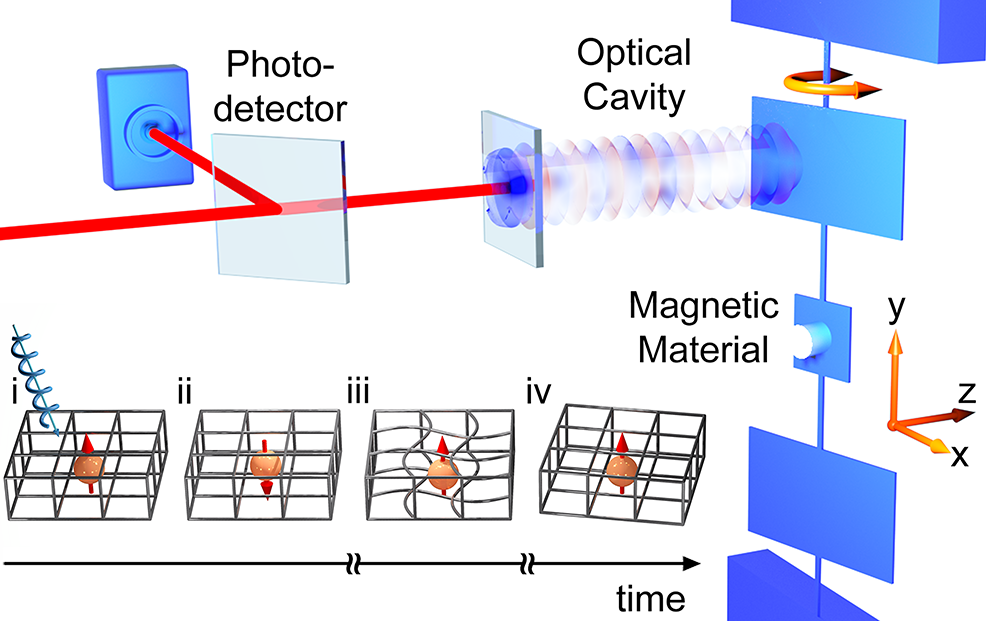}
	\caption{Schematic of a nano-optomechanical magnetic torque experiment.  The three wire-frame blocks along the notional time line suggest the microscopic sequence of events: i,ii) a source of magnetic torque/angular momentum pumping incident on an equilibrium spin in a lattice; iii) an instant of spin-lattice relaxation, conserving angular momentum by inducing a local twisting within the lattice (after Jaafar \emph{et al.} 2009 \cite{Jaafar2009}); iv) resultant rigid body rotation of the entire lattice, enabled by the specimen being placed on a soft torsional suspension.  The whole-body mechanical rotation is detected through sensitive optical interferometry.}  
	\label{Fig1}
\end{figure*}

Ongoing advances in the miniaturization and sensing of mechanical devices have set forth stimulating and exciting explorations into a vast range of scientific fundamental inquiries and applications such as atomic mass sensing \cite{Yang2006, Roy2018}, observation and control of quantum phenomena \cite{Teufel2009, OConnell2010}, metrology \cite{Wilson2015, Purdy2017}, and imminent photonics-based \cite{Bernier2017, Rath2013} devices.  A subset of these investigations are seeking a further understanding of small-scale magnetism through the detection of mechanical torques transduced via magnetic spins, which contain direct information about the magnetization and related static and dynamic processes.  The scope and potential for micromechanical studies of ferromagnetism were reviewed insightfully and comprehensively by John Moreland in this journal 15 years ago \cite{Moreland2003}.  In retrospect, that moment in time was the threshold of the era of the `single quantum'.  Within two years of the Moreland review, single electron spin resonance measurements were accomplished using several different methods: mechanically, by magnetic resonance force microscopy at $T=1.6\thinspace$K in a strong field gradient of $2\times 10^5\thinspace$T/m \cite{Rugar2004}; optically, via a dipole-allowed transition between spin triplet states of a defect centre (nitrogen vacancy in diamond), which also revealed hyperfine coupling to a single $^{14}$N nuclear spin \cite{Jelezko2004}; and electrically, in transport through a field-effect silicon transistor (coupling to a paramagnetic spin in the oxide) \cite{Xiao2004}. Within less than a decade, observations of another long-sought goal, that of mechanical resonators in their quantum ground states and manipulation of single phonons, were achieved (again by multiple means, through coupling to small microwave \cite{OConnell2010} and optical \cite{Chan2011} cavities).  In cavity optomechanics, entangled states of single phonons with single photons can be created and manipulated \cite{Seif2018}.  Some of these advances help to fill the gaps between what was possibly previously in mechanical studies of magnetism, and the realm of single spins, as we shall describe with a particular emphasis on cavity optomechanics. 

In keeping with the stunning richness of microscopic physics within magnetic materials that has been uncovered over the past three decades, \cite{Stamps2014, Sander2017} observations of mechanical rotations are possible not only in response to torques applied by external magnetic fields (DC, \cite{Rowland1878} AC, \cite{Einstein1915} and RF \cite{Alzetta1967}) but have and/or can be performed to capture spin-polarized currents \cite{Johnson1985} flowing onto a sensor and relaxing within the device \cite{Zolfagharkhani2008}, inducing a torque by angular momentum transfer \cite{Slonczewski1996, Berger1996}.  Such measurements are extensible to magnons \cite{Chumak2015} and presumably even to phonons \cite{Jungfleisch2018}.  There has been recent conceptual progress around phonon angular momentum and its role in spin-phonon processes and the Einstein de-Haas effect \cite{Zhang2014, Garanin2015}.

An optomechanical transduction schematic of a mechanical torque sensor supporting a magnetic specimen is shown in Fig.~\ref{Fig1}.  A high-finesse optical cavity is formed between the mirror and paddle experiencing deflection due to a mechanical torsion mode.  The lower part of Fig.~\ref{Fig1} is an aid to visualizing the relevant microscopic processes, through a cartoon zooming in to the scale of individual spins within the magnetic material.  In panel i), an incident source of magnetic torque is about to interact with a spin.  Panel ii) shows a flipped spin a very short time later, after an angular momentum-conserving interaction not involving the lattice.  After some evolution time characteristic of the material, the spin relaxes to equilibrium and transfers angular momentum to the lattice.  Panel iii) shows the outcome of the relaxation process manifest instantaneously as a lattice vibration with non-zero angular momentum.  Finally, panel iv) illustrates the whole-body rotation resulting after the microscopic twisting `message' has been communicated throughout the object at the speed of sound.  In this highly schematic scenario the initial spin flip can arise from many sources, including the effect of a changing external magnetic field (inducing motion of a domain wall past the spin location in a ferromagnet, or representing the change in net statistical polarization in a paramagnet, or the absorption of electromagnetic radiation at a spin resonance in either); or a spin-spin interaction.  

The small spring constants and moments of inertia allowed by the scaling of mechanical resonator dimensions to the micro- and nanoscale offers enhanced torque sensitivity, which correlates to a high magnetic spin (moment) sensitivity.  The miniature mechanical `torque sensors' couple well to optical cavity transduction schemes, granting further gains to detection sensitivity.  The torque sensors also interface to and allow measurement of a wide variety of magnetic specimens from patterned thin films and nanoparticles to three-dimensional bulk-derived crystals, and their measurement is generally non-invasive. 

The ability to measure a single, small magnetic element offers insight into the effects of geometric confinement and fabrication-induced defects, both on the magnetic resonance profile and hysteresis. Torque sensing is broadband.  Both magnetostatics and magnetization dynamics can generally be investigated using the same sensor, including spin resonance and relaxation processes.  Micro- and nanoscale torque sensors can be tailored in geometry such that sensing frequencies coincide with available instrumentation bandwidth, and adapt well to frequency or time-domain measurement schemes.  Their high sensitivities also permit measurement in air and at room temperature, minimizing instrumental complexity.  

\section{Principles}

Micro- and nanomechanical systems can be treated largely within the framework of bulk elasticity theory, as presented in textbooks, \cite{Timoshenko1962, Cleland2003} and monographs \cite{Schmid2016}.  Some differences can arise in materials properties \cite{Xinxin2003}, on account of surface effects \cite{Yang2011}, and in the treatment of dissipation \cite{Wilson2008, Svitelskiy2009} adding interest to the topic.  The field has been in development for over half a century \cite{Newell1968}.

\subsection{Basic torque sensor mechanics}
Conceptually, a resonant torque sensor parses into two elements, the torsion spring and moment of inertia, and one can determine the square of the fundamental torsion resonance frequency from the ratio of an effective spring constant, $\kappa_{\rm eff}$, and an effective moment of inertia, $I_{\rm eff}$.   

\begin{figure}[h] 
	\centering
	\includegraphics[width=1\linewidth] {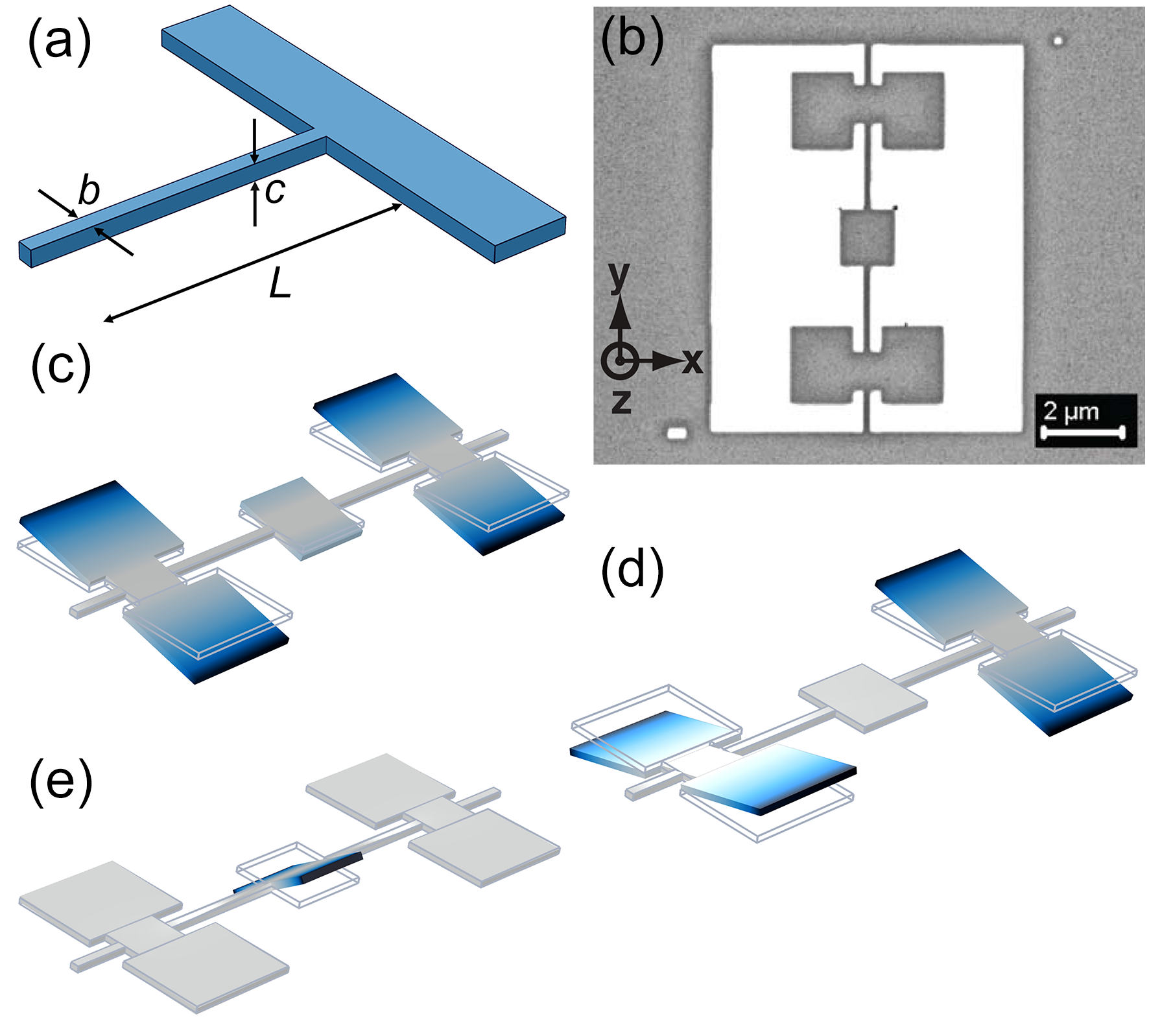}
	\caption{Basic torsion mechanics.  The simplest planar sensor geometry, a T-shape, is shown in a), indicating dimensions of the torsion beam.  b) A scanning electron micrograph of a triple paddle device patterned in a silicon nitride membrane by focussed ion beam milling (from Ref. [\citenum{Davis2010}], used with permission).  c,d,e) show the first three torsional eigenmodes as determined by finite element modelling \cite{COMSOL}.}  
	\label{Fig2}
\end{figure}

Fig.~\ref{Fig2}a illustrates a standard thin plate torsional resonator geometry, as can readily be manufactured from materials platforms including silicon nitride windows and silicon-on-insulator.  For design purposes the fundamental torsional resonance frequency may be estimated with standard elasticity theory, approximating the rod as the entirety of the torsion spring and the paddle as the sole contribution to moment of inertia.  Treating the torsion rod as a rectangular cross-section shaft with sides $b$, $c$, and length $L$, the spring constant is calculated from $\kappa = \beta G b c^3/L$, where $G$ is the shear modulus and $\beta$ is a geometrical factor correcting for the nonuniform shear stress around the perimeter (maximal at the middle of the long side; $\beta$ ranges between $1/3$ and $\sim 1/7$) \cite{BarredeSaintVenant1855}. Neglecting the moment of inertia of the spring in comparison to the paddle, we calculate the latter from $I = \int \rho_{\rm area} dA r^2$, approximating the paddles as thin rigid plates with areal mass density $\rho_{\rm area}$.  The modal frequency is then given by $f = \sqrt{\kappa/ I}$.  The accuracy with which the frequency can be predicted from simple elasticity theory, or how much $\kappa_{\rm eff}$ and $I_{\rm eff}$ yield a better estimate of $f_0$ differ from these values, depends on how well the actual motion is approximated by a uniformly twisting torsion element loaded by a rigid pendulum bob.   

Such a t-shaped sensor can easily be used to monitor two components of torque by utilizing the fundamental flexural (cantilever) mode in addition to the torsion mode.  From a design perspective it is important to know where the cantilever frequencies will be, even for applications not intending to utilize those modes.  If the modes are too close the torsion mechanical noise floor will be affected by the cantilever mode, and the modes will have a hybrid character (displacement profile with non-zero flex, in addition to the twist), both of which should be accounted for in the sensor calibration (see Section \ref{TM_calibration}).  

A doubly-clamped torsion design preferentially stiffens the flexure modes relative to the torsion modes and is generally more robust, even essential, when using very thin silicon device layers or window membranes.  For a symmetric design, the fundamental torsion resonance frequency goes up by square root of two (two equivalent torsion springs acting in parallel), while the flexural frequency goes up by a power of two. Leading order corrections to account both for twist in the paddle and the moments of inertia of the beams have been worked out by Lobontiu \emph{et al.} \cite{Lobontiu2006}.

A characteristic triple-paddle nanofabricated torque sensor is illustrated in Fig.~\ref{Fig2}b.  The physical implementation shown in the scanning electron micrograph was patterned by focussed ion beam from a 100$\thinspace$nm thick silicon nitride membrane.  A 10$\thinspace$nm film of permalloy was deposited on one surface for magnetometry tests.  The mechanical response of this structure is approximated well by a parsing into four individual torsion springs symmetrically attached to three rigid masses.  From the symmetry of the design there are only two distinct torsion constants, which we label $\kappa_{\rm outer}, \kappa_{\rm inner}$, and two moments of inertia: $I_{\rm outer}$, $I_{\rm inner}$.  This is sufficient information to estimate the frequencies of the first three torsional resonances.  

In the lowest frequency mode, all three paddles resonate with uniform angular displacement and the twist is all concentrated in the two outer springs.  The fundamental modal frequency is then $f_1 = \frac{1}{2 \pi} \sqrt{\frac{\kappa_1}{I_1}}$ where $\kappa_1 = 2\kappa_{\rm outer}$ (the two springs act in parallel) and $I_1 = 2I_{\rm outer} + I_{\rm inner}$.  The second torsion mode is characterized by antisymmetric twist of the outer paddles while the inner paddles remain stationary.  The inner paddle is thereby effectively a clamping point for the inner torsion springs, and the frequency $f_2$ can be estimated with $I_2 = I_{\rm outer}$ and $\kappa_2 = \kappa_{\rm outer} + \kappa_{\rm inner}$.  In the third torsional mode of this structure, the inner paddle resonates while the outer paddles remain stationary: $I_3 = I_{\rm inner}$, and $\kappa_3 = 2 \kappa_{\rm inner}$.  Taking $G = 1.0 \times 10^{11}\thinspace {\rm N/m^2}$ and $\rho_{\rm area} = 4.3 \times 10^{-4}\thinspace {\rm kg/m^2}$ (the complete structure was 100$\thinspace$nm of nitride with 10$\thinspace$nm of permalloy on one side, and 2$\thinspace$nm of gold on the other to prevent charging from the ion beam; carbon anti-charging layers are used more recently) the numerical values found for the specific device in Fig.~\ref{Fig2}b are $\kappa_{\rm inner} = 1.0 \times 10^{-12}\thinspace {\rm N m /rad^2}$, $\kappa_{\rm outer} = 2.0 \times 10^{-12}\thinspace {\rm N m /rad^2}$, $I_{\rm inner} = 1.4 \times 10^{-28}\thinspace {\rm kg \, m^2}$, and $I_{\rm outer} = 4.3 \times 10^{-27}\thinspace {\rm kg\, m^2}$.  This leads to the calculated frequencies given in Table \ref{table:mode_freq}, presented along with the mode frequencies found from the experiment.  

\begin{table}
\centering
\renewcommand{\arraystretch}{1.5}
\begin{tabular}{|>{\centering\arraybackslash}p{.85cm}|| >{\centering\arraybackslash}p{2.3cm}| >{\centering\arraybackslash}p{2.3cm}| >{\centering\arraybackslash}p{2.3cm} | }

 \hline
 mode & calculated $f$ (MHz) & experimental $f$ (MHz) & finite element $f$ (MHz)\\
 \hline
 1 & 3.4 & 3.36 & 3.59\\
 2 & 4.2 & 4.2\footnote{\mbox{observed by piezo actuation}} & 4.36\\
 3 & 19 & 21 & 19.6\\
 \hline
\end{tabular}
\caption{Comparison of the calculated, experimentally measured, and finite element model simulated mechanical resonance frequencies of the first three modes of the triple paddle resonator shown in Fig.~\ref{Fig2}b. }
\label{table:mode_freq}
\end{table}
\noindent The close correspondence for the fundamental mode is coincidental.  These approximate analyses are not known for their precision, and even the 10\% difference between calculation and measurement for the third mode is very good.  The second mode was not observed in magnetic torque actuation: in uniform fields with equal magnetic moment on each large paddle, the antisymmetric mode doesn’t couple to the drive.  A mode was seen at 4.2$\thinspace$MHz with a piezo disk shaking the device, but unlike the magnetically-driven modes the torsional nature of this mode was not confirmed by scanned interferometric imaging, \cite{Davis2010} and a whole forest of flexural modes were observed on account of the resonator being patterned within a larger nitride window \cite{Norcada}.  

It has become routine to model numerically the motions of these sensors using finite element structural mechanics \cite{COMSOL}.  Snapshots from the numerical eigenmode determination are shown in the lower set of panels in Fig.~\ref{Fig2} (which support the approximations used for the analytical estimates), and the computationally-estimated frequencies are given in the last column of Table \ref{table:mode_freq}.  The most particular advantage of the finite element modelling in this context arises when twisting and flexing motions become hybridized, as can easily happen even for the fundamental vibrations in arbitrary geometries, and will arise inevitably in high-enough frequency eigenmodes.   

The torsion constant sets the scale of angular displacement to be expected for a given applied magnetic torque.  Two important factors determine the torque sensitivity of a given measurement: the quality factors of the torsion resonances (signals routinely are modulated on resonance to ring the displacement up to larger amplitude), and the noise floor of the measurement system (which should be designed to be limited by the thermally-driven Brownian motion of the device, known as thermomechanical noise, and not by other technical noise sources).  In recent years there have been great breakthroughs in the understanding of how to achieve low mechanical dissipation in micro- and nanomechanical resonators.  The key is ‘stress dilution’ to reduce intrinsic losses from thermoelastic damping and extrinsic losses from coupling vibration out of the resonator at the clamping points.  High stress silicon nitride drumheads were found to exhibit very high $Q$’s (on the order of one million) \cite{Verbridge2008}.  Subsequently, very elegant schemes drawing on phononic bandstructure engineering to ‘soft clamp’ the resonators have yielded ultrahigh $Q$’s (up to 100 million) even at room temperature \cite{Tsaturyan2017}.  Torque sensitivity, and in particular how to calibrate the sensitivity via well-resolved thermomechanical noise, is discussed in the next section. 

\subsection{Thermomechanical calibration} \label{TM_calibration}

The fluctuation-dissipation theorem describes stochastic mechanical motions driven by heat through the equipartition of energy $k_{\rm B}T/2$ to every quadratic potential or kinetic energy term in a modal analysis of the structure \cite{Callen1951}.  This sets an absolute scale against which the amplitude of optomechanically-detected motion may be calibrated, provided only that the temperature of the mode is known.  The procedure was developed thoroughly first for flexural modes \cite{Albrecht1991}, and later extended to torsional modes \cite{Losby2012}.  Provided the modes have a purely flexural or torsional nature, it is natural to equate $k_{\rm B}T/2$ to the potential energy at maximum displacement (where the kinetic energy is zero), thereby setting a scale to the amplitude of Brownian motion.  (For Doppler-based velocimetry detection, one would refer to the kinetic energy at minimum displacement.)  For coupled modes, the same procedure is often invoked, performing an integration over the mode profile in a finite element model, captured at the instant of maximum displacement at the detection point \cite{Eichenfield2009c,Hauer2013}.  A careful study of the partitioning of energy in coupled modes has been published recently \cite{Schwarz2016}.  A most exciting development within the topic of thermomechanical calibrations is the discovery that quantum correlations can be resolved beneath the thermomechanical noise background and used as a primary reference to self-calibrate the mode temperature (demonstrated in Ref. [\citenum{Purdy2017}] for a mode with $\hbar\Omega \sim 2000 k_{B}T$).  Another compelling prospect is to metrologically-calibrate torque sensitivity via impulses of angular momentum imparted by radiation, as can be envisioned for extensions of the type of work reported in Ref. [\citenum{He2016}].

\section{Cavity optomechanics} \label{optomech}

\subsection{Description of transduction}

The push for devices with increasing levels of sensitivity has resulted in the devices getting smaller and smaller. At the highest `routine' torque sensitivity levels today \cite{Wu2017,Kim2017}, minimum feature sizes of 100s of \si{\nano\meter} yield \si{\zepto\newton\meter\per\sqrt{\hertz}} sensitivity. Gains made due to sensor size reduction, however, come at the expense of the ease of transducing the mechanical motion of the sensor. Fortunately, the use of optical cavities in nano/micro mechanical sensors has allowed the sensitivity of these devices to reach unprecedented levels.

\begin{figure}[h] 
	\centering
	\includegraphics[width=1\linewidth] {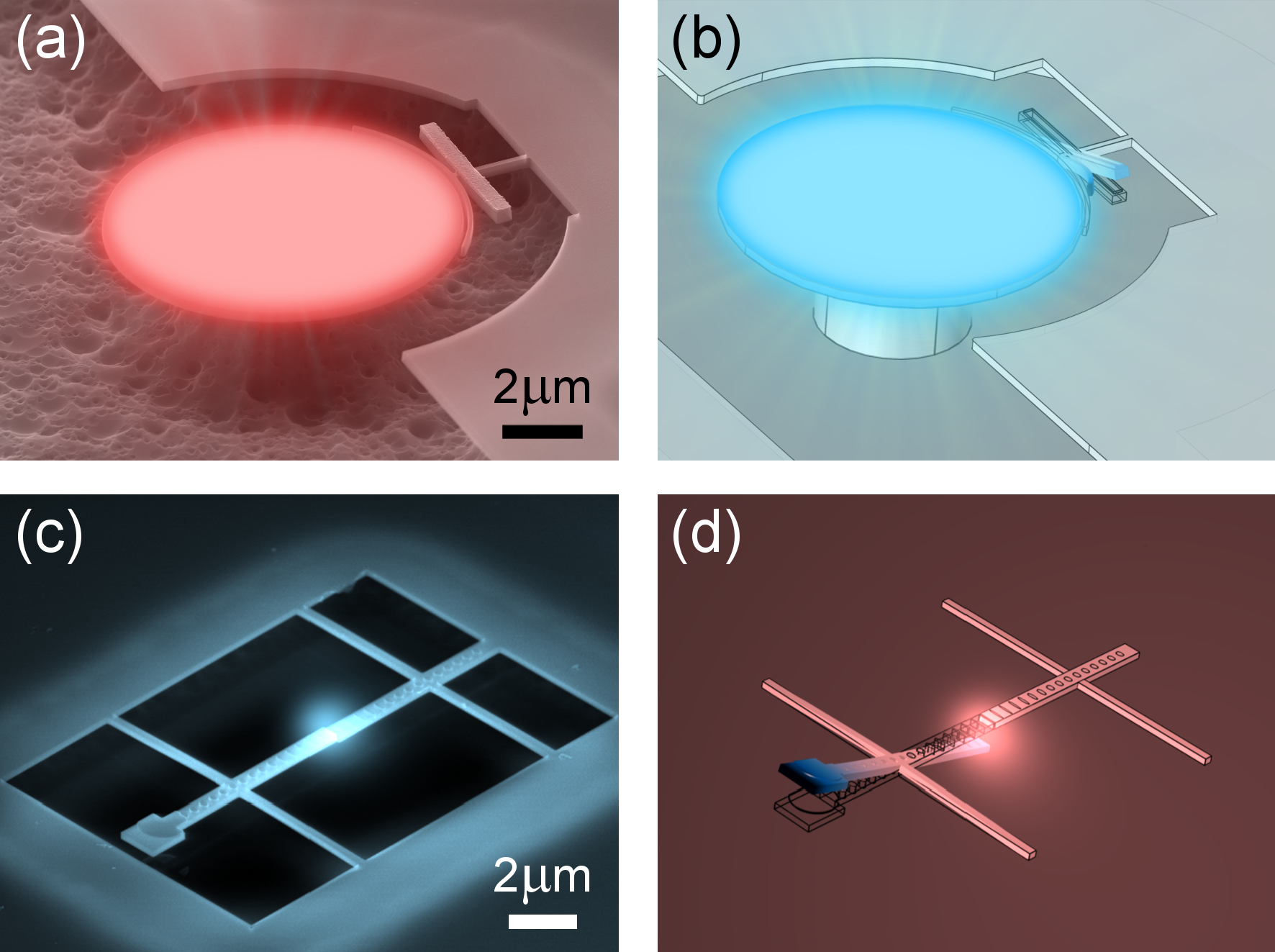}
	\caption{Optomechanical cavity transduction principle.  (a) Colourized SEM of an optical disk cavity with the mechanical resonator in a static equilibrium position. The optical power is enhanced in the disk. (b) Simulation of the exaggerated deflection of the torsional beam. The frequency within the cavity undergoes a blue shift as the local $n_\mathrm{eff}$ decreases with deflection. (c) Colourized SEM of a torsional split beam cavity device. (d) Simulation of the exaggerated deflection where the cavity will undergo a red shift with increasing deflection as the cavity becomes larger. (Figures (a,c) courtesy of P. Kim and J. Davis)  }  
	\label{fig:opticalcavitytransduction}
\end{figure}

In a cavity-optomechanical system, the mechanical motion of the sensor will most likely adjust the optical path length of the cavity (dispersive coupling). Other optical cavity variables that could change with the motion include the amount of light coupled into the system, or the optical losses present within the system (dissipative coupling) \cite{Elste2009,Wu2014}. The mechanical transduction is brought about through monitoring the optical power exiting the cavity. In the dispersive coupling case, which is the focus of this discussion, the optical path length changes and the characteristic optical cavity resonance frequency shifts accordingly, as illustrated in Fig.~\ref{fig:opticalcavitytransduction}. The oscillating optical resonance becomes a modulating output power through monitoring the cavity at a constant wavelength. The strongest signal occurs when the probe laser is set to the greatest slope of the optical resonance curve, typically a detuning off-resonance equal to half the optical linewidth.

Two main categories of optical cavities are used in conjunction with mechanical resonators: free-space and integrated. The free-space cavity is composed of two high reflectance mirrors having light coupled from free-space into the cavity to form a standing wave between the mirrors. In this implementation the mechanical resonator can interact with the cavity directly by selecting the mirror itself as the mechanical device \cite{Kleckner2006}, or it can be implemented through a membrane in the middle scheme in which a mechanical dielectric membrane is placed between two external mirrors \cite{Sankey2010, Reinhardt2016}. Alternatively, integrated optical cavities utilize on-chip integrated photonic elements to interact with a mechanical device. A typical device has the optical modes confined within the waveguide material (e.g. silicon, silicon oxide, silicon nitride, etc. -- foundry friendly materials) and the mechanical element is either the optical confining element itself \cite{SafaviNaeini2012}, or a separate entity which is coupled evanescently to the cavity \cite{Anetsberger2009,Li2009b}. Advantages of using integrated over free-space optics include the capability of using a much smaller optical mode volume more comparable in size to the mechanical device and the intrinsic alignment between optomechanical elements allowing for easier implementation. This can come at the expense of higher losses in the system due to the light propagating in lossy materials (the losses of which can follow non-linear processes which increase with higher fields) and fabrication-induced defects, both which can limit the optical finesse of the cavity system. Nevertheless, integrated optical cavities have been the most practical way to couple to mechanical devices on the sub-\si{\micro\meter} scale. 

This section will focus on the classical description of using cavity optomechanical systems to detect the motion of nanomechanical resonators. The analysis assumes dispersive coupling is dominant, and that the optomechanical system is in the Doppler regime (or unresolved-sideband regime where the mechanical frequency is much lower than the optical cavity decay rate). Here, it is noted that much of the current work using cavity optomechanics is geared toward studying quantum mechanical systems and occurs in the resolved-sideband regime where optomechanical cooling rates are much higher. A description of cavity quantum mechanical analysis is outside the scope of this review, but a thorough analysis by Aspelmeyer \emph{et al.}. can be found in reference [\citenum{Aspelmeyer2014}]. To begin, the cavity will be described mathematically. This description is similar regardless of whether the cavity is within a free-space or integrated system. Next, it will be shown how the motion of the mechanical element will perturb the cavity and how this can be measured. Effects including optical spring effects, and optical heating and cooling will also be discussed. 

The optical cavity can be described using a few key characteristic properties. The foremost is the optical frequency. This is determined by the whole number of wavelengths that can physically fit inside the cavity and is a factor of the dimensions and the effective index of refraction, $n_\mathrm{eff}$, of the optical mode, $m$. The effective index is analogous to the refractive index of light of a plane wave in free-space, but since light propagation in a waveguide has restricted transverse extension the index is not just a function of the material but also of the propagation mode and dimensions. Mode solving software can be used to calculate $n_\mathrm{eff}$ or it can estimated using the effective index method \cite{Kawano2001, Lifante2003}. For a cavity with a round trip length $L_{opt}$ the optical resonant frequency is 
\begin{equation}
\omega_\mathrm{opt,m} = m \frac{2\pi c}{n_\mathrm{eff} L_{opt}}.
\end{equation}
Like any resonator, it will have an associated bandwidth which is dictated by the losses in the device. These losses can be grouped into the total cavity decay rate, $\kappa_\mathrm{opt,m}$, which is the decay rate for the energy of the optical mode in the system. As such, the optical quality factor, $Q_\mathrm{opt}$ is defined as
\begin{equation} \label{eq:Q_opt}
Q_\mathrm{opt} = \frac{\omega_\mathrm{opt,m}}{\kappa_\mathrm{opt,m}}.
\end{equation}
The next characteristic property that will be discussed here is the free spectral range, $\FSR$, which is the spacing between two successive optical modes. Expressed in terms of optical frequency (or wavelength, following from the approximation $\delta \omega  /\omega = \delta \lambda / \lambda$), the $\FSR$ is \cite{Bogaerts2012} 
\begin{subequations}
\begin{align}
\FSR_\mathrm{\omega} &= \omega_\mathrm{opt,m+1} - \omega_\mathrm{opt,m} = \frac{2\pi c}{n_\mathrm{g} L_{opt}},\\
\FSR_\mathrm{\lambda} &= \frac{\lambda^2}{n_\mathrm{g} L_{opt}}.
\end{align}
\end{subequations}
The group index, $n_\mathrm{g}$, describes the effective group velocity of the light envelope ($n_\mathrm{eff}$ describes the phase velocity), and is used to take into account dispersion in the waveguide. The $\FSR$ is used to calculate the finesse of the cavity, $\mathcal{F}$, which gives the power enhancement of light in the cavity as it represents the average number of times the light circulates within it. The finesse is calculated as follows:
\begin{equation} \label{eq:finesse}
\mathcal{F} = \frac{\FSR_\mathrm{\omega}}{\kappa_\mathrm{opt,m}}.
\end{equation}
In equations \ref{eq:Q_opt} and \ref{eq:finesse}, $\kappa_\mathrm{opt,m}$ can be approximated by the full width half max, $\FWHM$, of the optical transmission curve when $Q_\mathrm{opt} \gg 1$ such that $\mathcal{F} = \FSR_{\omega(\lambda)} / \FWHM_{\omega(\lambda)}$. 

To determine the effect of the oscillating mechanical element on the optical cavity, the transmission of light through the cavity is described next. For simplicity, it will be assumed from here on out that the analysis is for a single optical cavity mode, so the subscript `m' can be dropped. Also, a rotating reference frame coinciding with the laser probe frequency, $\omega_\mathrm{p}$, is used such that the field $E_\mathrm{initial} = \exp(-i \omega_\mathrm{p}t) E$, and use $\Delta = \omega_\mathrm{p} - \omega_\mathrm{cav}$ as the laser detuning  compared to the cavity frequency, $\omega_\mathrm{cav}$.  

The analysis continues for the case of the unidirectionally coupled optical cavity. This is a common implementation for an integrated optical cavity, and corresponds to a system with a single input and single output of interest. Concrete examples include all-pass ring resonators or whispering gallery mode resonators, in which a single waveguide couples evanescently into and out of the adjacent cavity.   For an all-pass ring implementation, the transmission, $T_{opt}$, a ratio of the output optical power to input optical power in the cavity is \cite{Aspelmeyer2014}
\begin{equation} \label{eq:cav_transmission}
T_{opt} = \frac{\left ( \frac{\kappa_0 - \kappa_\mathrm{ex}}{2} \right )^2+\Delta^2}{\left ( \frac{\kappa_0 + \kappa_\mathrm{ex}}{2} \right )^2+\Delta^2}
\end{equation}
The total loss rate, $\kappa_\mathrm{opt}$, has been separated into individual contributions such that $\kappa_\mathrm{opt} = \kappa_\mathrm{ex} + \kappa_0$, representing the input coupling loss and the remaining cavity loss rate, respectively. From equation \ref{eq:cav_transmission} we can differentiate three different regimes of the cavity system. The first is the overcoupled system where coupling losses dominate ($\kappa_\mathrm{ex} \gg \kappa_0$). Conversely, an undercoupled system sees the internal cavity losses dominate. Lastly, there is the critically coupled regime in which the losses are equal ($\kappa_0 = \kappa_\mathrm{ex}$) and the cavity transmission equals zero at cavity resonance. This means that maximum energy is stored in the device when the input laser frequency matches the cavity resonance corresponding to the highest finesse. 

It can be difficult to design optomechanical systems with physically separate mechanical and optical elements while maintaining low optical losses, but nevertheless cavity detected nanomechanical sensors have achieved state-of-the-art sensitivity levels with modest optical $Q$ factors below 10000 \cite{Roy2018,Wu2017,Kim2017}. These larger damping rates bring the system into the Doppler regime, where $\kappa_\mathrm{opt} \gg \Omega_\mathrm{mech}$ and $\Omega_\mathrm{mech}$ is the resonant frequency of the mechanical device. 

This is a convenient regime for mechanical sensor systems as dynamic backaction effects (i.e. optical spring effects or optomechanical damping/antidamping on the mechanical device) can be minimized. Therefore, a gain in displacement sensitivity can be achieved without necessarily affecting the properties of the mechanical sensor. If those mechanical properties are the metric of detection, it removes any ambiguity that may arise as to whether or not the sensing action also affects the optical cavity properties and convolutes the measurement. 

With minimal backaction effects and small device displacements, the analysis of the mechanical device transduction is straight forward. It simplifies to a linear phase change in the cavity caused by the mechanical device's motion. With the phase, $\phi$, and mechanical displacement, $z$; it is expressed as follows \cite{Sauer2014, Diao2017}:
\begin{equation} \label{eq:transduction}
\frac{\partial T_{opt}}{\partial z} = \frac{\partial T_{opt}}{\partial \phi}\frac{\partial \phi}{\partial z}.
\end{equation}

This can be expanded using experimental parameters such as the slope of the transmission curve at the probe wavelength, $\partial T_{opt}/\partial \lambda$, and the effective index change due to the motion of the mechanical beam $\partial n_\mathrm{eff}/\partial z$. The latter can be simulated through effective mode index calculations. For a mechanical beam  with an optical cavity interaction length, $L_mech$:
\begin{equation} \label{eq:transduction_coefficient}
\frac{\partial T_{opt}}{\partial z} = \left [  \frac{\partial T_{opt}}{\partial \lambda}  \frac{\FSR_\lambda}{2 \pi}\right ]\left [  \frac{2 \pi}{\lambda_\mathrm{p}} \frac{\partial n_\mathrm{eff}}{\partial z}  L_\mathrm{mech}\right ].
\end{equation}

The first term on the right hand side of equation~\ref{eq:transduction} corresponds to the cavity alone, while the second term relates to the optomechanical coupling strength. For the highest transduction sensitivity, $\partial T_{opt}/\partial z$, one must maximize the transmission phase sensitivity, $\partial T_{opt}/\partial \phi$. This occurs when the probe laser is set to the greatest slope of the optical transmission curve. For an all-pass ring configuration with minimal chromatic dispersion effects and an on-resonance transmission value of $T_0$ \cite{Roels2011}:
\begin{equation} \label{eq:maxdTdphi}
\left | \frac{\partial T_{opt}}{\partial \phi} \right |_\mathrm{max} \approx \frac{3 \sqrt{3}}{8 \pi} (1 - T_0) \mathcal{F}.
\end{equation}


Combining equation~\ref{eq:maxdTdphi} with equation~\ref{eq:transduction} and expressing it in terms of the optomechanical coupling constant, $g_\mathrm{om} = \partial \omega_\mathrm{opt}/\partial z$, and the cavity decay rate, the maximum transduction response for the minimal backaction and unresolved sideband regime is
\begin{equation}
\left | \frac{\partial T_{opt}}{\partial z} \right |_\mathrm{max} \approx  \frac{3 \sqrt{3}}{8 \pi} (1 - T_0)  \frac{2\pi}{\kappa_\mathrm{opt}} g_\mathrm{om} .
\end{equation}
It is important in the above equation to account only for the local index change caused by the mechanical beam on the whole cavity such that $|g_\mathrm{om}|=2\pi c \, (n_\mathrm{eff}L_{opt})^{-2} \,(\partial n_\mathrm{eff}/\partial z)\, L$. 

\subsection{Comparison with other methods}

It is interesting to compare the techniques of optical transduction: free-space interferometery, integrated non-cavity interferometry, and integrated cavity transduction.  Here, free-space interferometry refers to a typical Fabry-Perot nanomechanical beam detection setup \cite{Carr1998, Ekinci2005}, where a low-finesse cavity is formed between the nanomechanical device and the back substrate (usually spaced a distance dictated by the sacrificial release layer thickness of the nanomechanical device: $\sim$100s of \si{\nano\meter}). This detection method is compared directly to integrated racetrack
\begin{figure}[h] 
	\centering
	\includegraphics[width=1\linewidth] {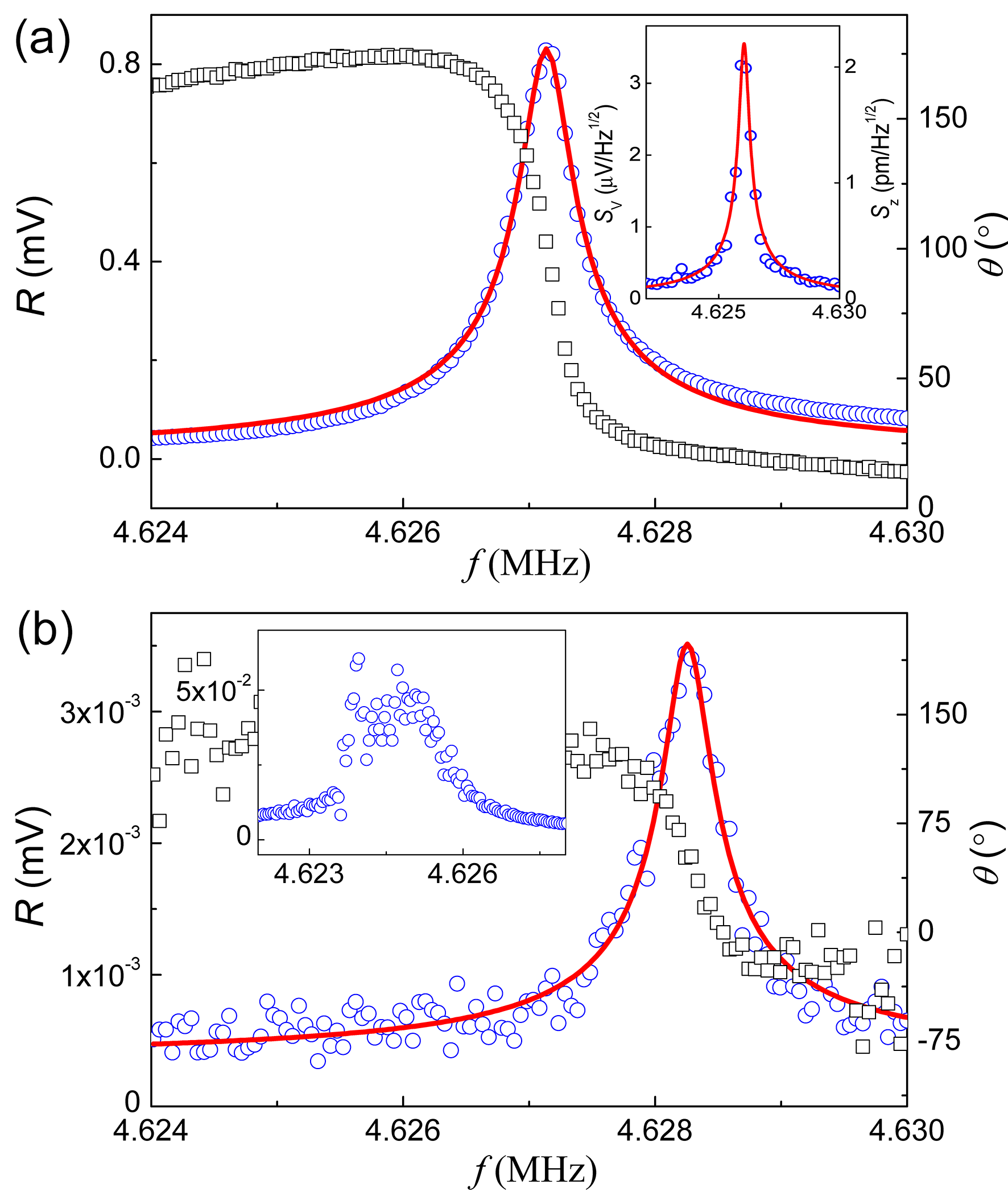}
	\caption{Comparison of the mechanical response of a \SI{15}{\micro\meter} long, \SI{500}{\nano\meter} wide, and \SI{220}{\nano\meter} thick doubly clamped beam being detected by (a) a racetrack resonator optical cavity (RTR) with \SI{47}{\micro\watt} laser power at the device and (b) a free-space Fabry-Perot setup with \SI{28}{\micro\watt} laser power at the device. The inset of (a) is the detected thermomechanical response and the inset of (b) is the free-space response with \SI{80}{\micro\watt} laser power.  (from Ref.  [\citenum{Diao2013a}], used with permission.)}  
	\label{fig:freespace_vs_cavity}
\end{figure}
resonator cavity transduction by probing the same doubly clamped beam (integrated within the racetrack resonator cavity) using both methods \cite{Diao2013a}. The signal is enhanced two orders of magnitude while also avoiding the pitfalls of the Fabry-Perot method for detecting nanomechanical structures. These include the limitations set by the diffraction limit when detecting sub-micron sized structures and also the stability required when aligning to such small structures. This enhancement of the signal is shown in Fig.~\ref{fig:freespace_vs_cavity} where the cavity detection in (a) is orders of magnitude higher compared to the free-space signal in (b). Most importantly, the thermomechanical noise of the signal is resolvable with the cavity transduction scheme, but not with the free-space signal. 

\begin{figure}[h] 
	\centering
	\includegraphics[width=1\linewidth] {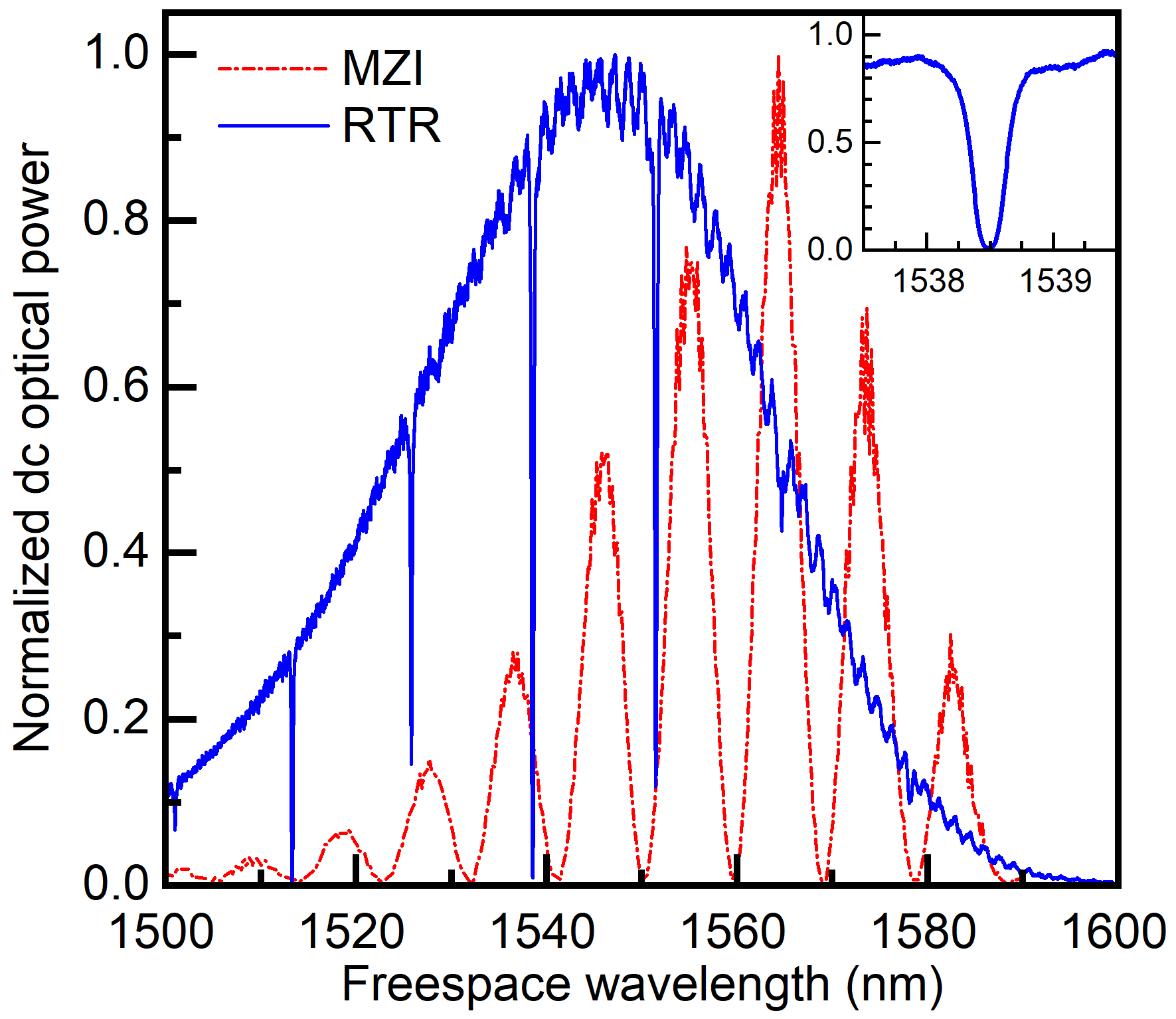}
	\caption{Comparison of the dc optical transmission curves of a non-cavity Mach-Zehnder Interferrometer (MZI) and racetrack resonator optical cavity (RTR) device. The inset shows a single cavity resonance dip for the RTR device.}  
	\label{MZIvsRTR}
\end{figure}

Transduction enhancements also occur between two similar nanomechanical cantilever devices both measured using integrated nanophotonics, one using a Mach-Zehnder interferometer implementation and the other using a racetrack resonator optical cavity \cite{Sauer2014, Sauer2012}. The mechanical devices are independent of the waveguide elements to remove any size restrictions waveguide criteria would place on them. When used as a sensor, the separate optical and mechanical components also allow the mechanical sensor to contain metal features without causing significant optical losses due to unwanted plasmonic interactions \cite{Kim2017}. The $\partial T_\mathrm{opt}/\partial \lambda$ at the probe location is directly proportional to the signal, and plotting their normalized dc optical transmission curves in Fig.~\ref{MZIvsRTR} illustrates the drastic difference in their responsivities. This coupled with the increased phase accumulation due to the cavity's finesse allows a three order of magnitude improvement in the displacement sensitivity (non-cavity: $\sim$\si{\pico\meter\per\sqrt{\hertz}}; cavity: $\sim$\si{\femto\meter\per\sqrt{\hertz}}).
The sensitivity enhancement provided by optomechanics has unleashed new opportunities for studying fundamental physical properties. At one end, using strongly coupled optomechanical systems allow the quantum regime of micro-scale mechanical objects to be accessed. At the other end, using weakly coupled optomechanical systems provides such significant enhancement in the displacement sensitivity of a nanomechanical device that its thermomechanical noise becomes the limiting noise source - even at room temperature and atmospheric pressure \cite{Roy2018,Wu2017}. Although design and fabrication restrictions may make it difficult to implement strongly coupled optomechanical systems in all experiments, the sensitivity improvements (even in modest systems) make it an enticing technique.

\section{Nanofabrication of hybrid magneto-mechanical systems} \label{nanofabrication}

Advances in miniature mechanical sensing and micromagnetism have been developing alongside those in nanofabrication, where feature sizes down to the tens of nanometers are now possible.  This progress has transpired in the form of improving traditional methods and the invention of new techniques.  Photolithography, developed in the 1960s during the rise of the integrated circuit architectures, is still employed as resolutions are pushed to the nanometer regime with the use of optical wavelengths into the deep and extreme ultraviolet, as well as through projection stepping methods.  Electron beam lithography (EBL), originally performed using modified scanning electron microscopes, are dedicated systems capable of sub-10$\thinspace$nm exposure line-widths.  Other fabrication techniques of mechanical structures involve the use of focussed ion beam milling as well as 3D microprinting.  We concentrate here on the nanofabrication of highly sensitive optomechanical sensors that can be driven by magnetic torques.  These comprise hybrid nanomechanical resonators with magnetic specimens affixed to them (through a multitude of means), or mechanical devices created fully from magnetic materials.  

\subsection{Fabrication of torque sensors} 
\begin{figure*}[ht] 
	\centering
	\includegraphics[width=1\linewidth] {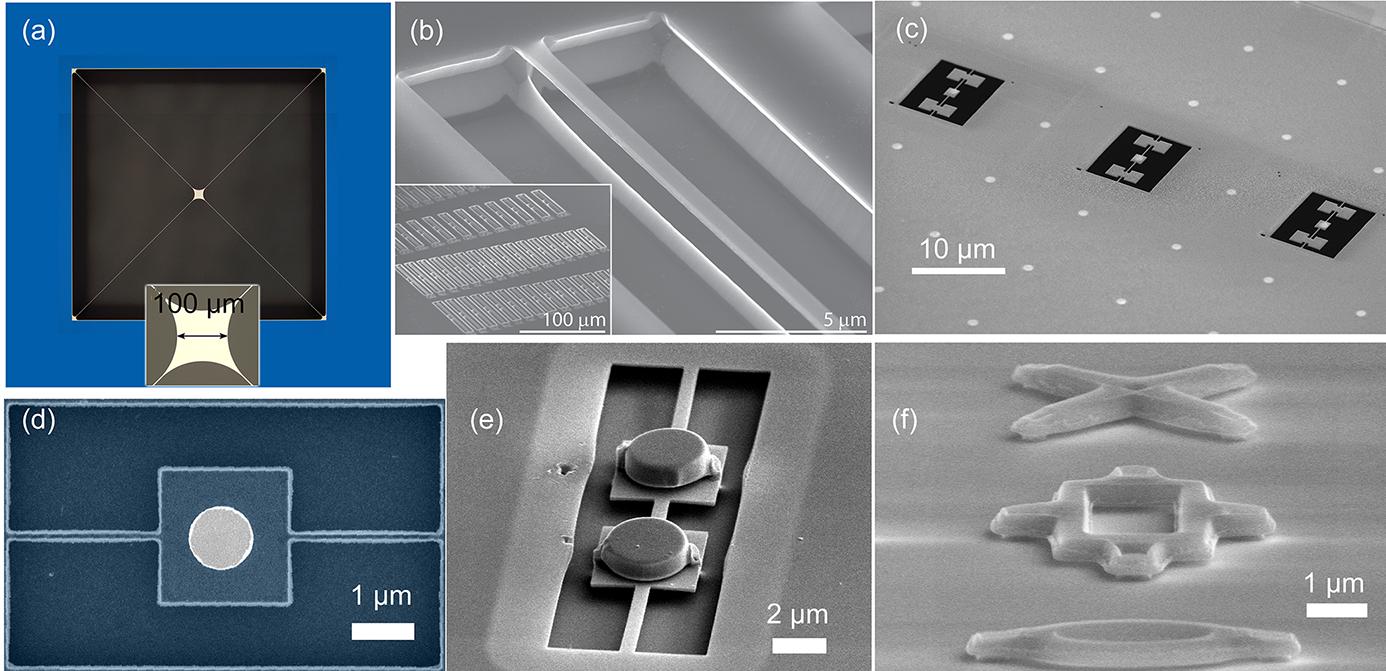}
	\caption{Mechanical resonators and hybrid magnetic torque sensors. a) a trampoline structure (thickness 80$\thinspace$nm at center) inset: magnification of center region (from Ref.[\citenum{Reinhardt2016}], used with permission). b) Single-crystal diamond nanoribbon with the fabricated array shown in the inset (from Ref.[\citenum{Khanaliloo2015}], used with permission).   c) Three triple paddle torque sensors defined by focussed ion milling around pre-patterned with permalloy disks on a silicon nitride membrane. d) Silicon-on-insulator-based torsional paddle holding a 30$\thinspace$nm thick cobalt (20$\thinspace$nm)/cobalt oxide (10$\thinspace$nm) disk. e) Double-paddle torque sensor supporting two yttrium iron garnet microdisks which were milled by focussed ion beam from a bulk crystal, and nanomanipulated onto the sensor. f) Free-standing YIG structures fabricated with a photo-resist templated GGG substrate (from Ref. [\citenum{Heyroth2018}], used with permission).}  
	\label{Fig3}
\end{figure*}
The first consideration in designing a nanomechanical torque sensor is its, as well as the magnetic material's, geometry with respect to applied magnetic fields.  The magnetic torque expression ($\tau = M \times  \mu_0 H$), requires that the cross-product of the magnetic moment and an applied field induces a magnetic torque in a direction orthogonal to both.  This torque should be oriented to drive a compliant mechanical mode.  For a torsional mode, as shown in the schematic of Fig.~\ref{Fig1}, the torque generated by net magnetization in the $\hat{x}$ direction by a field in the $\hat{z}$ direction is directed along the torsion bar ($\hat{y}$ direction).  For the measurement of direct torque, it is ideal to `dither' the $z$ field at the mechanical resonance frequency for maximum sensitivity.  Magnetic torques can also be transferred to cantilever or doubly-clamped beam flexural modes, where the torque direction is perpendicular to the mechanical beam length axis and applied field.  A sweepable DC bias field is provided along the magnetization axis for magnetic hysteresis loop characterizations.  Applied field modulation and demodulation schemes for driving and detection magnetic resonance are also possible with the introduction of other AC field components and directions, and discussed in Section \ref{spin_resonance}.  

The silicon-on-insulator (SOI) architecture has proven to be a popular precursor material for the fabrication of nanoscale mechanical resonators, optical resonators and photonic cavities, as well as optical waveguides.  Engineered to reduce current leakage in integrated circuits, SOI substrates consist of an insulator (a silicon oxide) sandwiched in between a thin silicon `device' layer and a thick silicon `handle' layer.  The mechanical structure is defined in the device layer through optical or electron beam lithography patterning with an appropriate polymer resist, which after development provides an etch mask for transferring the pattern into the silicon.  The device layer is often etched via reactive ions, in which the ions generated from a plasma in an RF field are accelerated towards the sample.  Anisotropic etching of the silicon is achieved by physical and chemical bombardment from the ions, alternating with a passivation process that coats the trench sidewalls to inhibit lateral etching.  After the remaining resist is stripped, the silicon oxide layer is etched (in a hydroflouric acid based solution or vapor, which selects only the oxide) in order to release the structure.  

The thicknesses of the device and handle layers can be tuned depending on the detection scheme.  In the case of Fabry-Perot type interferometric free-space optical detection, the ideal silicon oxide and device layer thicknesses should offer maximum constructive interference contrast of the reflected intensity through the cycle of the mechanical oscillation.  For optical cavity and waveguide based detection methods, the device layer is often tuned to about a 220$\thinspace$nm thickness to allow for optimum localization of single-mode propagating photons, usually at telecommunication wavelengths (as discussed in the Appendix). 

Thin silicon nitride membranes can be acquired commercially with varying stresses and thicknesses (down to 10$\thinspace$nm), and are a very popular choice as the starting material for nanomechanical systems \cite{Norcada}.  The definition of mechanical sensors within a membrane can be done lithographically or directly through focussed ion milling, as presented in Fig.~\ref{Fig2}b.  Stresses in the membranes, as well as their thinness, can present challenges to their survival through fabrication processes.  In lithographic processes, care must be taken during resist spinning as well as wet etching (home-built carriers, for example, have been made to allow flow of etching solution while holding the membranes rigidly in place \cite{Reinhardt2016}).  A silicon nitride based `trampoline' resonator is shown in Fig.~\ref{Fig3}a. For focussed ion beam milling of membranes, dosages and milling geometries should be tuned in order to prevent heat and ion-damage induced curling, especially in high stress regions.  


The dawn of single-crystal diamond optomechanics has paved a way for idealized devices with remarkable mechanical, optical, and thermal properties.  Owing to it's large bandgap ($\sim$5.47$\thinspace$eV, with multiphoton absorption bands in between ~2500$\thinspace$nm and $\sim$6700$\thinspace$nm), diamond in the IR and optical regimes have a maximum theoretical transmission on the order of 70$\%$, allowing for high optical power confinement \cite{Coe2000}.  The optical properties combined with high stiffness, low mechanical dissipation, and high thermal conductivity yield ideal cooperativities in optomechanical coupling. The bulk diamond substrates, commercially available, are grown through homoepitaxial chemical vapor deposition, where the seed layer is synthetic or naturally occurring diamond \cite{Balmer2009,Yan2002}.  The difficulties of top-down fabrication of diamond micro- and nanomechanical resonators arise from not only their difficulty in etching, but also having adequate lithographic masking that can survive the process.  In recent developments, silicon nitride and silica based hard masks have proven to be effective for enduring several hours of exposure to the ICP-RIE \cite{Tao2014, Khanaliloo2015, Mouradian2017, Ovartchaiyapong2012}.  Integration of single-crystal diamond devices with resonant optical cavities is at an exciting stage.  Shown in Fig.~\ref{Fig3}b is a diamond `nanoribbon' which can be coupled to an optical cavity through nearby tapered optical fibre.


\subsection{Hybrid sample preparation} 

Affixing microscopic magnets to nanomechanical torque sensors presents several challenges.  These arise from magnetic materials not being able to withstand most wet chemical and plasma etching processes (while retaining magnetic `purity') in silicon-based lithography as well as the difficulty of controllably attaching magnetic material to a suspended nanomechanical element (without destroying the small structure or causing its stiction to substrate surfaces).  These issues have recently been addressed through several approaches, with much success.  

One such method involves direct focussed ion beam (FIB) milling of a torque sensor holding a micromagnetic disk in a Si$_{3}$N$_{4}$ membrane.  Prior to ion beam definition, an array of permalloy (Ni$_{80}$Fe$_{20}$) disks were deposited in ultrahigh vacuum with a collimated electron beam though a shadow mask of holes defined in another Si$_{3}$N$_{4}$ membrane.  With careful alignment (usually to features defined outside the milling region), the sensor geometry is cut into the membrane from a software pattern (which allows for the parameterizing of beam milling directions and dosages).  In order to prevent overexposure of the magnetic material to gallium ions, the FIB milling of the torque sensor is done `bottom side up', where the disks are on the Si$_{3}$N$_{4}$ membrane opposite to the side facing the incident beam.  Three triple paddle torque sensors milled in this fashion into a membrane are shown in Fig.~\ref{Fig3}c.  A similar technique was also employed to define sensors around magnetic nanoparticles deposited on Si$_{3}$N$_{4}$ membranes \cite{Firdous2015}. These devices can be integrated into or as optical cavities, where the Si$_{3}$N$_{4}$ membrane can be attached (with adequate spacing) to a mirror.  

In the case of multi-layer and SOI-based lithography, it is often preferable to attach the magnetic material at or near the end of the fabrication process to prevent exposure to etching steps that can affect its surface quality (where notches, pits, and other geometric defects induce magnetization pinning sites).  Magnetic thin disks have been deposited onto pre-fabricated nanomechanical torque sensors by manual alignment of micrometer sized holes defined in a Si$_{3}$N$_{4}$ membrane (commercially available as transmission electron microscope specimen holders) acting as a shadow mask \cite{Losby2012}, though difficulty in alignment and destruction of sensors during the process will affect yield.  A lithographic step developed to grant higher accuracy in placement of confined thin film structures onto mechanical sensors, through EBL definition of the patterns onto a photoresist spun on the suspended structures.  A key step is avoiding the stiction due to nanoscale surface forces that can occur if the space between the device layer (in the case of an SOI sandwich) and substrate layer is thin, and the suspended structure is affected in a way to cause it to collapse onto the substrate.  For the case of photoresist spinning onto the sensor the device chip is immersed in an intermediary and fast-evaporating solvent, which supports the suspended structure, while slowly being replaced by flowing photoresist \cite {Diao2013}.  Standard EBL procedure, thin film deposition, and liftoff followed by critical point drying allows for the creation of varied magnetic geometries with placement accuracy within 50 to 100$\thinspace$nm (limited by how well alignment marks are observed under the photoresist during EBL).  A torque sensor housing a layered micromagnetic disk (20$\thinspace$nm cobalt, 10$\thinspace$nm cobalt oxide) created using this process is shown in Fig.~\ref{Fig3}d.           
Some magnetic materials are difficult to incorporate into traditional lithographic processes because they cannot be directly grown, evaporated, or deposited on the mechanical sensor.  These include magnetic crystals epitaxially grown on a seed layer of matching atomic lattice parameter, of which a popular class for micromagnetic applications are the magnetic garnets, namely yttrium iron garnet (YIG, Y$_3$Fe$_5$O$_{12}$).  Due to difficulty of forming nanoscale confined magnetic geometries with YIG through its growth,    a focussed ion beam was used to mill a micromagnetic shape into a bulk crystal, which was `plucked and placed' onto a torque sensor using a nanomanipulator \cite{Losby2015}.  The micromagnet is often `tacked' onto the torque sensor through electron- or focussed ion beam-assisted deposition of material at their interface.  A double-paddle sensor geometry housing two YIG disks attached using this method is shown in Fig.~\ref{Fig3}e.  Consideration of redeposition and, more importantly, the creation of a thin `dead layer' encapsulating the magnetic volume due to ion damage is necessary when adjusting milling geometries and parameters.  The `pluck and place' method can be applied to shape and transfer a variety of bulk-derived small magnetic materials.  Torque sensors that are fully comprised of garnets are also desirable, and have been accomplished using focussed ion beams \cite{Fraser2010, Seo2017}.  Free standing YIG structures have also been cleverly grown on templated GGG surfaces, and offers promise as a method to form torque sensors not affected by ion milling, as shown in Fig.~\ref{Fig3}f \cite{Heyroth2018}. 

\section{Recent torque studies}

\subsection{Magnetic hysteresis, quantifying the Barkhausen effect} \label{magnetic_hysteresis}

The low radio frequencies (MHz) of operation of typical nanomechanical torque sensors make them more immune to laboratory sources of noise (and in particular $1/f$ noise) than competing methods of measuring equilibrium hysteresis curves of small magnets, such as magneto-optics.  The long-term sub-microradian polarization stability required to resolve (for example) changes of order one part in a thousand of the total moment is by comparison much more difficult to achieve.  In contrast, exquisitely-resolved stroboscopic magneto-optical measurements of fast magnetization dynamics are comparatively easier to achieve \cite{Chen2012}.  It is interesting to note that in the vast majority of such stroboscopic studies, the equilibrium hysteresis of the magnetic structure under investigation has not been characterized (particularly when the measurement is performed on a single object as opposed to an array, exploiting the ability of magneto-optics to have good coupling to a small structure through focussing).  The very good agreements between experiment and micromagnetic simulations generally found rely in part on driving the system sufficiently away from equilibrium that the system can be approximated as ideal for the purposes of modeling. 
\begin{figure*}[ht] 
	\centering
	\includegraphics[width=1\linewidth] {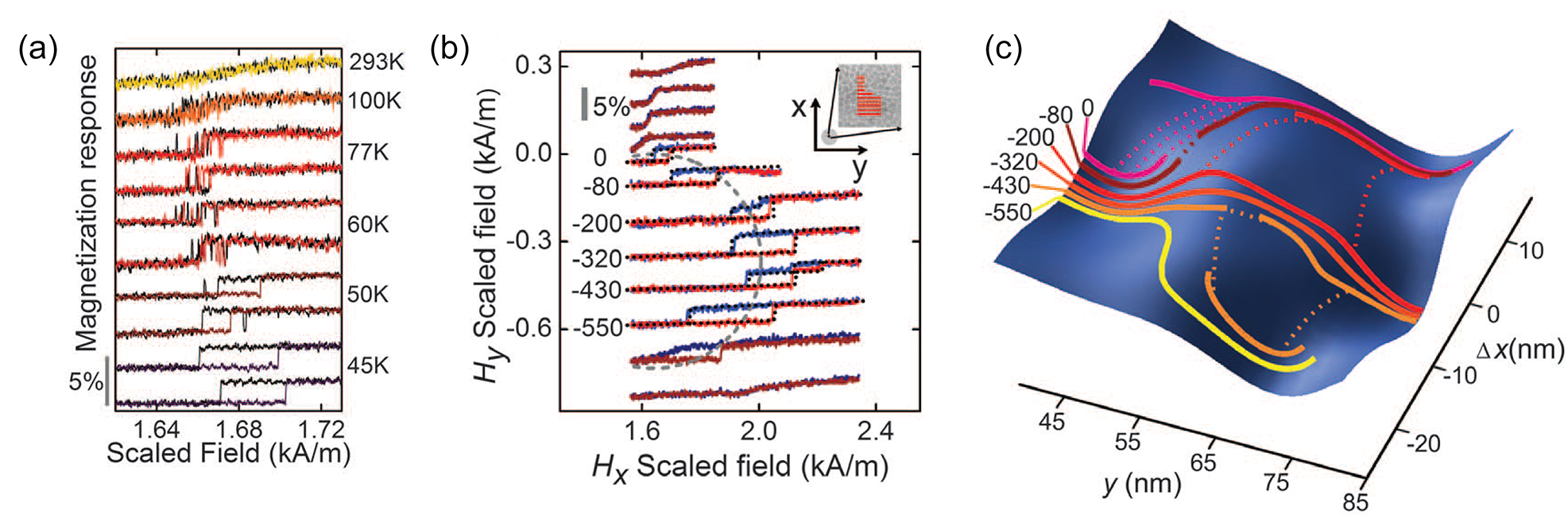}
	\caption{Quantification of the Barkhausen effect through torque magnetometry.  a) The evolving signature of thermally-activated vortex core hopping between a specific pair of neighbouring pinning sites in a 1 $\thinspace \mu$m diameter, 42$\thinspace$nm thick permalloy disk, tracked as a function of temperature.  Each trace is acquired at the same speed, scanning the applied field up and back across this narrow range.  At room temperature, the hopping rate is much higher than the measurement bandwidth and no hysteresis is observed: at each field value, the torque signal represents the time average of the net moment at the two positions, weighted by the relative dwell times in the two pinning sites.  At liquid nitrogen temperatures, telegraph noise is observed as the bias field sweeps through the range of the minimum energy barrier between the two sites (the field applies a tilt to the overall energy landscape).  Finally, at low enough temperatures, the only signature of thermal activation is small variation from scan to scan of the field positions of the single hopping transition in each direction of the minor hysteresis loop (no additional telegraph noise).  The scaled field plotted as the x-axis corrects for the temperature dependence of saturation magnetization, $M_\mathrm{s}$.  The magnetic field change required in order to push the vortex core a given distance in the disk is proportional to $M_\mathrm{s}$.   b) Mapping minor hysteresis loops over a fine grid of $H_x$ and $H_y$ captures many subtle features as the vortex core finds its way in the 2D disorder potential underlying the Barkhausen effect.  c) The 2D energy landscape as determined by reduction of the hysteresis loop data in b), using an analytical model of vortex state energetics \cite{Burgess2014}.  The solid lines are minimum energy trajectories, while the dashed lines connect approximate jumping-off and landing points of vortex core hops.  (Figures from Ref. [\citenum{Burgess2013}], used with permission.)}
	\label{Fig6dot1}
\end{figure*}
An example of how mechanical torque measurements can assist in elucidating otherwise obscure details of small-scale magnetic phenomena is found in quantification of the energetics of domain walls in materials with nanoscale inhomogeneities of magnetic properties (saturation magnetization, exchange strength, anisotropy) for example at grain boundaries, or on account of surface or edge roughness and overall confinement geometry.  The exquisite sensitivity of a mechanically-resonant nanoscale torque sensor can open sufficient measurement bandwidth to monitor slow dynamics such as thermally-activated hopping of high energy density topological structures in the spin texture (domain walls) between adjacent locations (pinning sites), where the energy of the structure is locally reduced on account of the material inhomogeneity.  An example of this is shown in Fig.~\ref{Fig6dot1}a.  Temporal recordings of telegraph noise can map the exponentially-sensitive hopping rates as a function of external applied field and temperature, yielding quantitative determination of energy barriers.

In the specific instance of a vortex core in a soft ferromagnet, the high spatial localization of the topological spin structure can form the basis of a raster-scanned microscopy of the energy landscape.  Fig.~\ref{Fig6dot1}b shows the Barkhausen features in net $m_x$ for a mapping over an $(H_x,H_y)$ field range approximately 1$\thinspace$kA/m on a side, including bistability (`forward' and 'backward' $H_x$ scan directions).  The fine details (Barkhausen features) in the magnetizing curves can be analyzed to reveal a 2D energy map \cite{Burgess2013}.  When the vortex core resides in a small circular disk, an analytical model of the single-vortex spin texture greatly assists in the transformation from magnetic moment to energy.  The model accounts for the change of moment both as a function of vortex core position and from flexing of the (springy!) texture in the rest of the disk while the more remains pinned as the external field is changed, and when the core depins \cite{Burgess2014}.  The result from analysis of the data in Fig.~\ref{Fig6dot1}b is shown in Fig.~\ref{Fig6dot1}c.  Variations with position of the vortex texture energy in the disk on the scale of 100$\thinspace$meV are seen over a spatial patch mapped by the core position about 30$\thinspace$nm in extent in both in-plane directions.    

Another advantage for stroboscopic magneto-optics in the context of nonequilibrium dynamics is that higher optical powers can be used than for equilibrium measurements.  A reflected probe pulse carries away the polarization information of interest {\em before} the absorbed heat from that pulse perturbs the specimen.  The heating may be neglected, for repetitive phenomena, provided one waits long enough for the sample to cool again before the next pulse sequence arrives (often, $\sim \mu$s delay).  

Overall, and at the present time, magneto-optics and nanomechanical torque measurements are highly complementary, and best tailored for fast dynamics (magneto-optics) and equilibrium or slow dynamics measurements (nanomechanics).  For equilibrium magneto-optical studies, the unavoidable direct, continuous laser heating is problematic for measurements on small structures.  Combining stroboscopic magneto-optics with torque measurements, simultaneously on the same specimen, has been attempted but presents additional challenges for cooling the specimen between laser pulses.  (In vacuum, the only channel for heat to escape is by conduction through the torsion bar.)  Future combinations of nanomechanics with full-frame stroboscopic X-ray magnetic imaging could be more promising, for example with specimens supported on torque sensors fabricated from X-ray and electron transparent thin membranes \cite{Stohr1998}.  The next section describes the current state-of-the-art for all-mechanical simultaneous combined measurements of spin statics and spin dynamics.

\subsection{Spin resonance} \label{spin_resonance}

In magnetostatic or equilibrium measurements, we often think of the magnetic torque in terms of the first derivative of the magnetic potential energy as a function of angular position, as described in Section \ref{nanofabrication}. This leads to helpful descriptions for understanding the strengths of torques arising from magnetic anisotropies (magnetocrystalline and shape), for example.  If a specimen is fully isotropic, it is not possible to subject it to a static torque through application of a steady magnetic field — its magnetic moment can move into alignment with the field without any need for rotation of the mechanical object `hosting' that magnetic moment.  Its spins are decoupled from its lattice, as it were.  This is the crucial reason why it is important for compass needles to be needle-shaped: to hold their moment direction within the physical object \footnote{within a factor of $\pi$ radians; the moment direction can be reversed through judicious or unintentional application of transient or strong fields, causing the compass to read backwards}, such that the magnetic torque on the spins is transferred completely into mechanical torque on the object.  Much of micro- and nanomechanical torque magnetometry amounts to miniaturization of the principle of the compass.  

The description just given is incomplete, however.  When a magnetic field not parallel to the spins is applied suddenly to a magnetic system, conservation of angular momentum requires (in a classical description) the moment to precess around the new resultant field direction rather than rotate directly toward it; correspondingly, a quantum spin system will undergo Rabi oscillations.  Physically, this occurs because magnetic dipoles intrinsically carry mechanical angular momentum.  A time-rate-of-change of this mechanical angular momentum also can give rise to a net mechanical torque on the physical object hosting those spins (known as an Einstein-de Haas effect; see Section \ref{Einstein_deHass}).  

An ingenious application of this principle to magnetic resonance was conceived by Gozzini \cite{Gozzini1964} and first implemented by Alzetta, Ascoli {\em et al.} in 1967 \cite{Alzetta1967}.  In their demonstration, the electron spin resonance in an electron paramagnetic organic free radical solid was driven continuously save for amplitude modulation at the low torsional resonance frequency of a torsion fibre pendulum the specimen was suspended from.  With even a weak DC field applied along the torsion axis, creating a small net statistical polarization of the spin ensemble, angular momentum absorbed from the radio frequency magnetic field driving the spin resonance transfers to the mechanical host through spin-lattice relaxation yielding a measurable mechanical displacement, with the modulation giving rise to a resonant torque on the pendulum.  In the initial work of Alzetta {\em et al.}, the spin-lattice relaxation time, $T_1$, of their paramagnet (DPPH) was about 100$\thinspace$ns, roughly seven orders of magnitude shorter than the torsion period of the pendulum.  By pumping the paramagnetic resonance near saturation one has the small number of spins represented by the statistical net polarization undergoing approximately $10^7$ cycles of angular momentum absorption from RF radiation followed by angular momentum transfer to the lattice, within each period of pendular angular motion.  This beautiful technique has been surprisingly under-utilized, even despite a much later re-demonstration in a micromechanical format by the same group following the introduction of magnetic force microscopy \cite{Martin1987, Saenz1987} and magnetic resonance force microscopy \cite{Evans1956, Sidles1991, Rugar1992}.  

The pioneering work of Gozzini and co-workers constitutes the first mechanical torque-based study of nonequilibrium (nonequilibrium steady-state) spin physics.  The only shortcoming of this approach from the perspective of a putative all-mechanical tool for concurrent observations of spin statics (hysteresis) and spin dynamics (resonances) in micro- and nanomagnets is that the DC bias field is along the torsion axis; if the net moment of the specimen is in that direction, it cannot yield a measurable equilibrium torque.  
\begin{figure}[h] 
	\centering
	\includegraphics[width=1\linewidth] {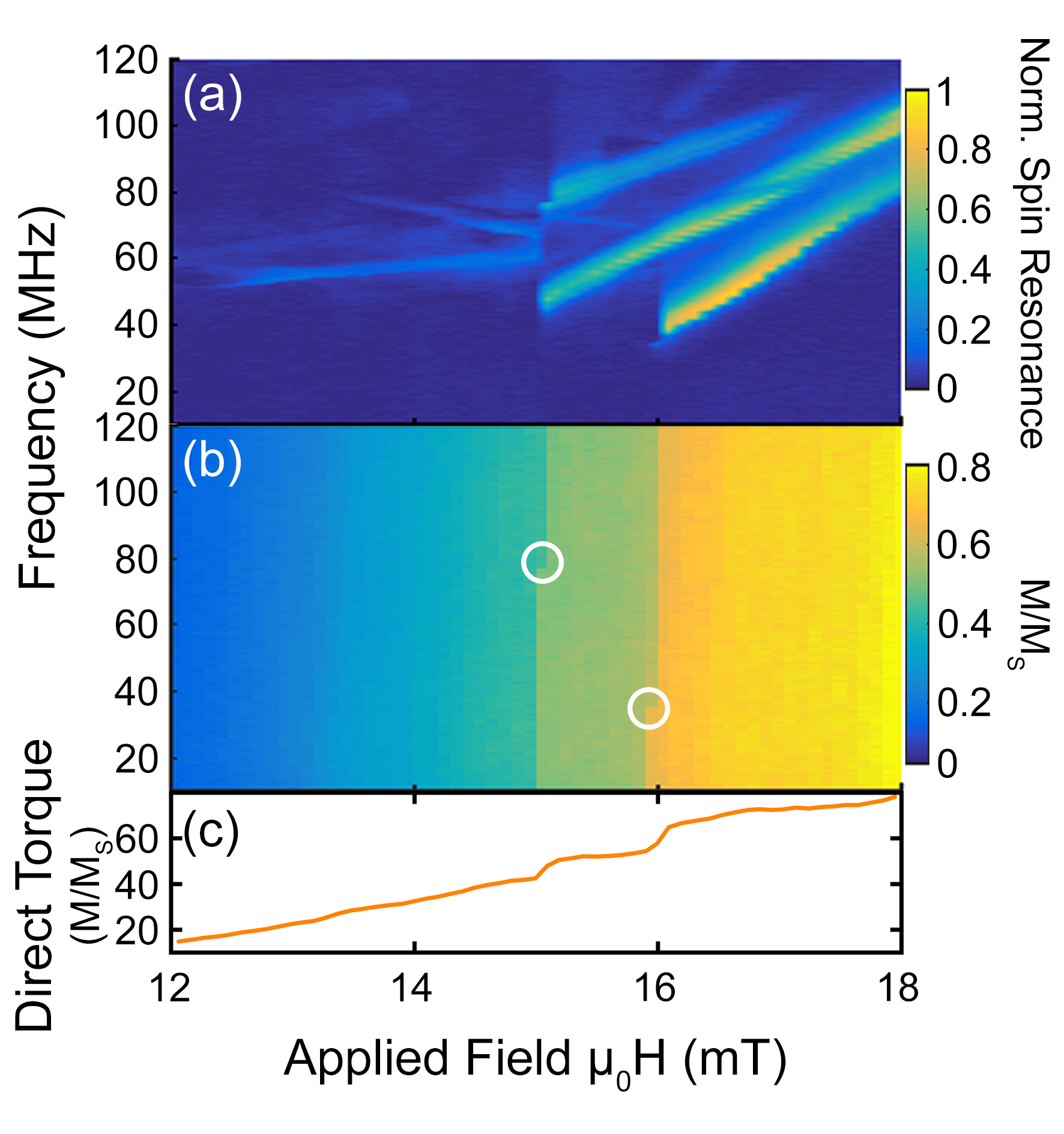}
	\caption{Simultaneous magnetometry and spin resonance spectroscopy.  The sample here consists of two yttrium iron garnet microdisks, mounted one per paddle on a double-paddle torque nanomechanical sensor (as shown in Fig.~\ref{Fig3}e).  The applied field is stepped down (high to low), and a resonance spectrum is acquired at each field step by sweeping the magnetic resonance drive and torque-mixing frequencies from low to high.  Panel (a) maps the resonance signatures in the RF susceptibility, and the simultaneously-recorded net moment (direct torque) is shown in (b).  Small circles indicate a resonance-assisted vortex nucleation transition in each disk, where the resonance modes of the high field spin texture abruptly vanish, correlated with a discrete downward jump of the net moment. (c) Net moment from (b) averaged along the field axis indicating two nucleation processes.}
	\label{Fig6dot2}
\end{figure}
This field geometry constraint can be lifted by an alternative approach to choreographing the magnetic resonance information to yield a mechanically-resonant torque on the sensor.  An approach to this is illustrated in Fig.~\ref{Fig6dot2}.  Beginning in the usual magnetometry configuration with a DC field to bias the specimen perpendicular to the torsion rod (along $\hat{z}$ in the adopted coordinates), consider a magnetic resonance driven by an RF excitation $H_1$ along $\hat{y}$, perpendicular to the DC field and parallel to the torsion rod.  Visualizing the magnetic resonance response classically, RF components of magnetic moment arise along $\hat{x}$ and $\hat{y}$ from the magnetization precession pumped by the RF drive.  Applying an additional small RF magnetic field tone $H_2$, detuned from the drive tone by $f_{\rm mech}$ and along $\hat{z}$, will induce a mechanically-resonant torque at $f_{\rm mech}$ along $\hat{y}$ by mixing (sum and difference frequency generation in the multiplication of the two terms).  In the limit of small driving fields this signal will be linearly proportional to both RF field strengths, $H_1$ and $H_2$.  In addition to nonlinear responses to larger $H_1$ (scales set by saturation in the case of paramagnetic resonances, and instabilities to shorter wavelength spin wave generation for ferromagnetic resonances), for linearity to $H_2$ amplitude must be small relative to the field line width of the magnetic mode because $H_2$ modulates $H_0$.  (Note, however, that unlike a similar modulation at much lower frequency as might be used for lockin detection of a signal, for example in a conventional inductive measurement of EPR, the modulation here is {\em not} taking the first derivative of the lineshape.) 

Continuing with the classical description, the $H_2$ tone may be viewed in the rotating frame (where one circularly-polarized component of $H_1$ is stationary) as an AC magnetic field along $\hat{z}$ at frequency $f_{\rm mech}$.  On account of $H_1$ torquing the moment away from equilibrium, the precession driven by $H_2$ contributes an Einstein-de Haas-like AC torque on the sensor.  There are no constraints on the mechanical frequency relative to the magnetic resonance frequency or linewidth in order for this method to function. 


Observation of a mechanical torque is always predicated on spin-lattice relaxation, however, and the signal should disappear when the mechanical frequency too far exceeds the relaxation rate.  One of the most interesting research directions for torque-mixing magnetic resonance spectroscopy (TMRS) is developing means for measuring spin-lattice lifetimes.  Detecting resonances with this approach is not a greater technical challenge than inductive measurements, it is merely different.  Just as inductive spectroscopy would be of much less utility without capabilities to measure lifetimes, more comprehensive capabilities of TMRS need to be fleshed out. 

In application to ferro- and ferrimagnetic resonances, many more field geometries will give rise to a resonance condition.  Driving a nonuniform spin texture into resonance will in general yield RF moment components in all Cartesian directions.  This can be captured in a simple susceptibility tensor description with off-diagonal terms.  The simultaneous magnetometry and spectroscopy enabled through all-mechanical detection allows direct correlation of resonance modes with the demarcations of different textures indicated by transitions in hysteresis loops.  An example is shown in Fig.~\ref{Fig6dot2}.  The sample here consists of two YIG disks on a double-paddle torque sensor (one disk per paddle), similar to one shown in Fig.~\ref{Fig3}e.  The two disks have slightly different vortex nucleation fields on the upper branch of the hysteresis loop (field sweep direction high-to-low; in the measurement, the frequency was swept low-to-high at each field step).  Clearly visible in the resonance spectra are the pixels at which vortex nucleation occurred in each disk.  Characteristically in such measurements, \cite{Losby2015} the nucleations are found to be resonance-assisted, usually occurring as the frequency scan begins to climb up onto a magnetic resonance mode.


\subsection{Micromechanical Einstein-de Haas studies} \label{Einstein_deHass}

In 1915 Einstein and de Haas published the first results from their gyromagnetic investigations of the origin of magnetism in materials.  They succeeded in observing a mechanical rotation induced by changing magnetization that had eluded earlier investigators \cite{Maxwell1954, Richardson1908}, although their initial determination of the gyromagnetic ratio famously proved erroneous by approximately a factor of -2.  Peter Galison has written an exceptionally detailed and insightful history of the early Einstein-de Haas and Barnett effect experiments \cite{Galison1987}.  Between Maxwell and Einstein-de Haas’s times a firm prediction for the scale of the effects had emerged thanks to the discovery of the electron and independent determinations of the electron charge-to-mass ratio.  The smallness of the anticipated effect meant that the elimination of measurement artifacts would consume the major part of the experimentalist’s attention.  Years of iteration in multiple laboratories were required before the discrepancy between measurements and the molecular current model of magnetism became undisputed, followed by decades more research to obtain precision measurements elucidating materials-dependence of the magnetomechanical factor, $g \prime$, the observed ratio of magnetic moment to mechanical angular momentum \cite{Scott1962, vanVleck1951, Kittel1949}.
\begin{figure}[h] 
	\centering
	\includegraphics[width=1\linewidth] {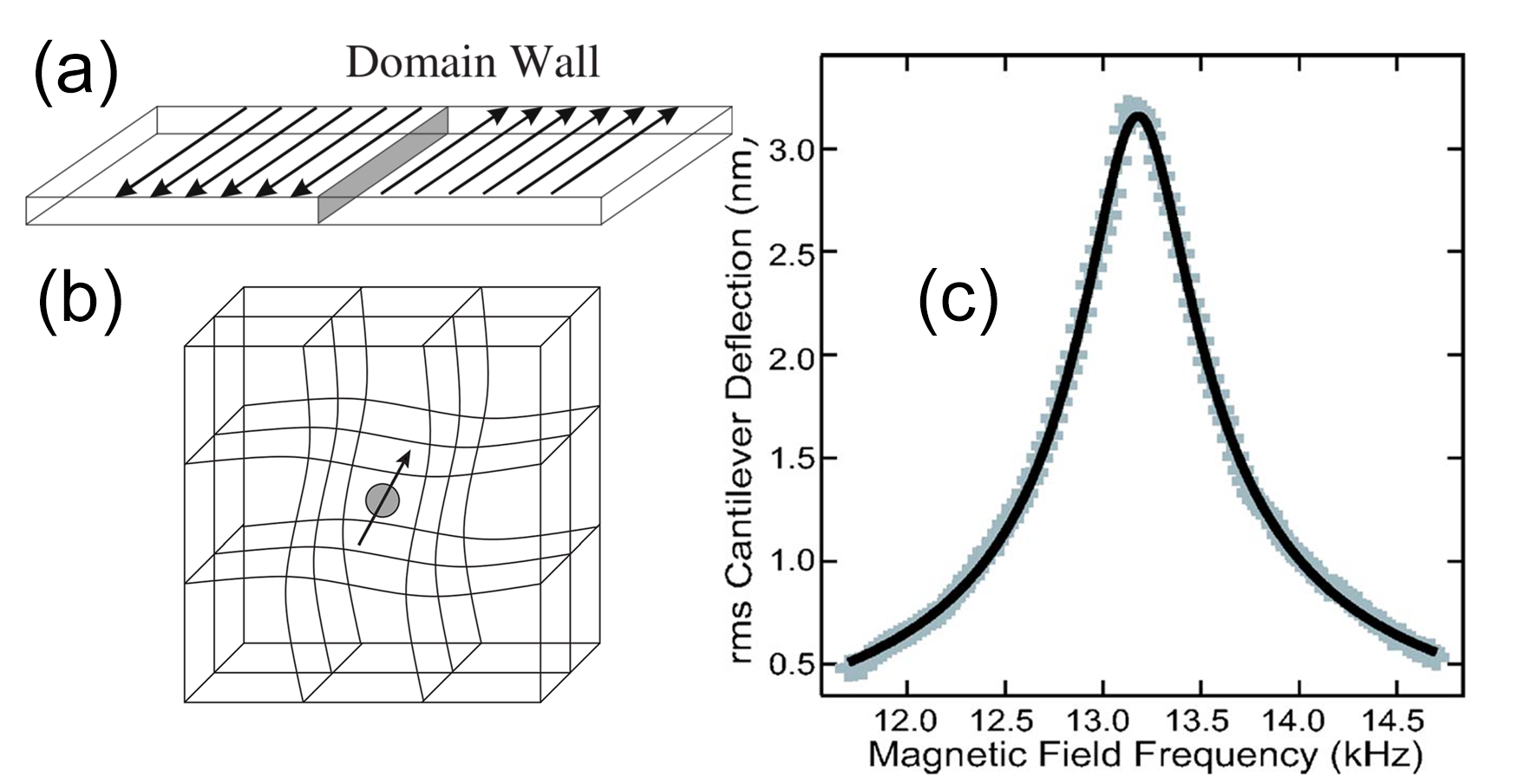}
	\caption{Micromechanical Einstein-de Haas measurement of the magnetomechanical ratio of permalloy, using a thin film deposited on a cantilever.  a) The experiment is described by a model in which the AC magnetic susceptibility is dictated by the reversible motion of a single domain wall within the film (provided the AC field strength is sufficient to overcome any domain wall pinning).  The data, (c), show the mechanically-resonant signal enhancement as the AC drive frequency is swept through the fundamental cantilever resonance (the flexural mode is driven by Einstein-de Haas torque along $\hat{x}$). The wire frame cartoon, (b), suggests the instant of microscopic transferrance of angular momentum from a spin to the lattice.  (Figures from Refs.[\citenum{Jaafar2009},\citenum{Wallis2006}], used with permission.)}
	\label{Fig6dot3}
\end{figure}
Micro- and nanomechanical implementations of Einstein-de Haas style measurements offer several clear advantages and disadvantages relative to the earlier, more macroscopic implementations.  The sample and sensor necessarily are still part of a macroscopic apparatus, but with the magnetic material under investigation much more localized it is easier to apply uniform magnetic fields and avoid background effects.  The higher frequencies of small devices for mechanically-resonant detection play a special role, as not only do they aid in decoupling from mechanical noise sources but the increase of net torque from the same repetitive magnetic moment/angular momentum change at higher frequency can compensate for the higher stiffness of small sensors.  The main disadvantage is the mundane difficulty of integrating the material of interest with the small sensor.  As the advantages have accrued in disproportion to the disadvantages, we are witnessing a slow resurgence of the study of Einstein-de Haas effects in small magnetic structures.  The pioneering micro-mechanical measurement of a magneto-mechanical $g \prime$ was performed by Wallis \emph{et al.} \cite{Wallis2006}, Fig.~\ref{Fig6dot3}.  The result is competitive with although not yet as precise as the earlier culmination of macroscopic measurements, after decades of refinement, described by Scott.  There remains a large space for iteration of the micromechanical measurements.  The small scale makes it easier to apply uniform magnetic fields.  With three-axis vector control of the fields, and a calibrated torque device with two modes sensitive to torques in orthogonal directions, in principle one can use a single specimen on a single device to determine the magnetic moment (correcting for anisotropy) and Einstein-de Haas torque, and hence $g \prime$.  Another very important consideration with micromagnetic specimens is the domain structure, and this will introduce systematic errors if it is not well-understood.  The Chudnovsky group has analyzed the Moreland-style cantilever-based Einstein-de Haas measurement of a ferromagnetic thin film on a cantilever under the assumption that the magnetization reverses via the motion of a single domain wall.   If the domain structure is well-characterized it can become an advantage. 

\subsection{Advances utilizing electromechanical detection (and non-detection!)}

A direct measurement of the angular momentum of light was reported by Beth in 1936 \cite{Beth1936}.  A small mechanical reaction on a waveplate, used to change the circular polarization state of light transmitted through it, could be measured by suspending the waveplate from a torsion fibre parallel to the optical axis.  A nanophotonic implementation of the Beth experiment has been performed recently by He and Li \cite{He2016}.  The mechanically-detected spin resonance experiments already described also rely on angular momentum transfer from (RF) photons.  


In an analogous experiment using spin-polarized electrons, Zohlfargakani \emph{et al.} in 2008 reported the direct observation of a mechanical angular momentum component in spin momentum transport \cite{Zolfagharkhani2008}.  A spin-polarized current was prepared off the device and transported through a long spin diffusion length of wire (cobalt) to an interface with a short spin diffusion length conductor (gold) near the centre of the device.  The gold depolarizes the spin current, acting as a spin momentum sink transferring net angular momentum to the lattice/body of the torque sensor.  This idea is extensible to other mechanisms of angular momentum transport, including those related to magnons \cite{Herring1951} and phonons \cite{Akhiezer1959}.  

The Zohlfargakani experiment was entirely electromechanical, and used capacitance-based displacement detection.  Commercial membrane stress sensors with piezoresistive readouts have been utilized very recently to demonstrate both force- and torque-detection of electron spin resonance \cite{Takahashi2018}.  

In a remarkable line of electrical measurements, electron transport through a suspended carbon nanotube, decorated by an individual single molecule magnet, is used to detect spin flips and even quantum tunnelling of magnetic moment, through inelastic electron scattering signatures in the current-voltage characteristic.  Among the most astonishing results of these studies is the {\em non}-observation of quantum tunnelling under circumstances in which the nanotube behaves like a mechanical vacuum for the molecule: when there is no mechanical mode to receive angular momentum from the flipping spin, the transition is precluded \cite{Ganzhorn2016}.

\subsection{Results from optomechanical implementations}

The great sensitivity enhancement derived from cavity optomechanics, as described in Section \ref{optomech}, is the single technical advance with the most promise for promoting chip-based mechanical probes of magnetism from niche/novelty applications to widespread use.  The full scope of applications is much broader than torque sensing and includes magnetic field strength magnetometers based on mechanical frequency shifts induced by magnetostriction \cite{Yu2016} and magnetic force microscopy (implicit in cavity sensors for atomic force microscopy) \cite{Srinivasan2011}.  Optical micro- and nanocavity enhanced torque sensors have been implemented recently, coupling nanomechanical torsional resonators either to whispering gallery mode microdisk resonators or to 1D photonic crystal cavities.  

\subsubsection{Measurements with whispering gallery mode microdisk readout of nanomechanical torque sensors}

\begin{figure}[h] 
	\centering
	\includegraphics[width=1\linewidth] {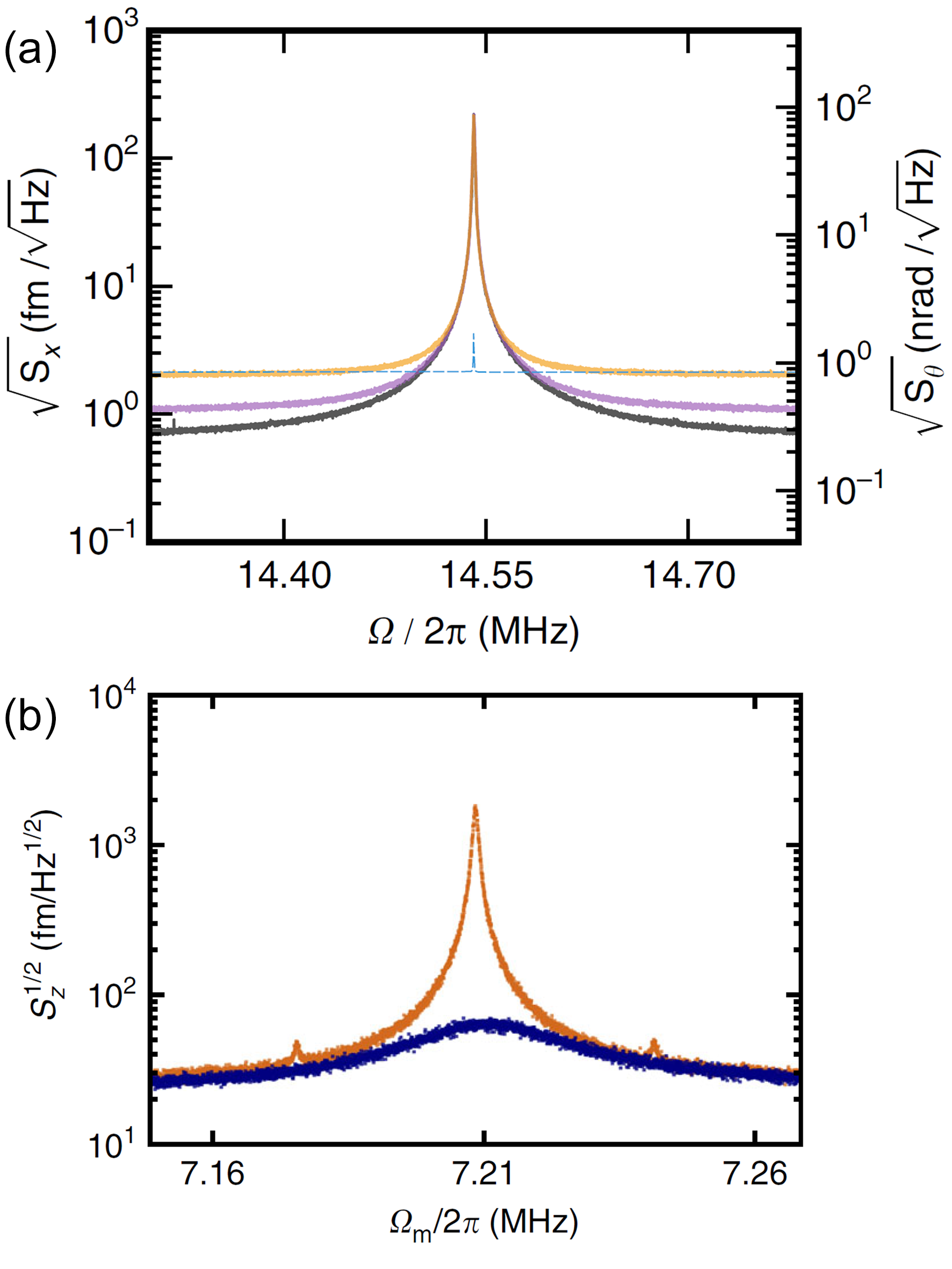}
	\caption{Thermomechanical displacement noise and torque sensitivity of microdisk optomechanical cavity readouts.  (a)  A measurement at $T = 4\thinspace$K, showing a very cleanly resolved thermomechanical peak and the improvement of displacement sensitivity with increasing optical power (data acquired in the regime where optical shot noise is the dominant source of uncertainty).  (b) Measurements on a different device performed at room temperature.  The orange trace shows the thermomechanical peak, while the blue trace (also acquired with the system at room temperature) demonstrates the ability to cool the torsional mode through magnetic torque feedback.  The feedback-cooled modal temperature as determined by fitting to the blue trace is $T= 12\thinspace$K. (Figures from Refs.[\citenum{Kim2016}, \citenum{Kim2017}], used with permission.)}
	\label{Fig6dot5one}
\end{figure}

The extant record for minimum resolvable torque in a 1 second measurement is held by a device with an arc-shaped paddle giving it a significant interaction length along the circumference of a microdisk \cite{Kim2016}.  With an additional central paddle to support the magnetic specimen of interest, this style of device, designed by Kim, Hauer and Davis, is known affectionately as `Pluto' for its resemblance to the sign of the zodiac (see Fig.~\ref{fig:opticalcavitytransduction}a and b).  By cooling the device in a dilution refrigerator to reduce thermomechanical noise (Fig.~\ref{Fig6dot5one}), a torque sensitivity of \SI{3e-24}{\newton\meter\per\sqrt{\hertz}} was achieved, albeit with a caveat: optical sampling had to be restricted to 20 ms light bursts to avoid degrading the sensitivity that results from the heating caused by the observation. Nevertheless this represents the first approach, to within an order of magnitude, of the standard quantum limit for torque sensing.  Note that, per the analyses from the early days of gravitational wave antenna design \cite{Braginsky1985}, time-windowed observations are of additional interest because the thermomechanical noise floor is not an intrinsic limit for measurements made within an inverse mechanical linewidth (coherence time of the mechanical mode).  In observations over a sufficient length of time, the thermomechanical background decreases through incoherent averaging as the inverse square root of the measurement time.  With high enough sensitivity, short observations, in contrast, can use the short-term phase predictability of the Brownian motion to resolve changes smaller than the thermomechanical amplitude.  
Operating at room temperature, the well-resolved thermomechanical signals from Pluto devices have been replayed back to the device in a negative magnetic torque feedback loop to reduce the Brownian amplitude (feedback cooling) \cite{Kim2017}. This is shown in Fig.~\ref{Fig6dot5one}b.  This approach does not change the minimum torque sensitivity for magnetic signals (since the feedback also acts on those signals), but it it does decrease mechanical ring-up/ring-down times and, hence, decreases the useful mechanical bandwidth for measurement (of transient response, for example).  Feedback cooling is another feature of optomechanical devices that extends their utility for applications \cite{Arcizet2006, Poggio2007}.  Another enhancement for which a proof-of-principle demonstration has been made, but that has remained otherwise unexploited for torsional devices, is parametric amplification \cite{Carr2000}.  

\subsubsection{Measurements with photonic crystal cavity readout}

Bragg reflectors in 1D photonic crystal waveguides can form the end mirrors of Fabry-Perot cavities having a total mode volume on the order of one cubic wavelength.  This extreme confinement of light is especially useful for sensing vibrations of very small nanomechanical resonators, with the marquee example being the optomechanical crystal, in which microwave frequency acoustic modes of the optical nanocavity itself are studied \cite{Eichenfield2009b}.  (Sometimes additional phononic band structure engineering is not even required to co-localize cavity phonons with the photons, and result in reasonably high mechanical $Q$s.)

The first application of optomechanical crystals integrated with nanomechanical torque sensors is in a design by Hryciw and Barclay \cite{Hryciw2013} incorporating an invention they call the split-beam cavity (SBC).  In an SBC, the Bragg mirrors are made by varying the ellipticity of the holes in the photonic crystal waveguide.  By elongating the holes sufficiently in the transverse direction, a cavity may be formed in which the centre (and highest electric field region) is a small gap at which the waveguide is cut all the way through, allowing one side to move relative to the other provided at least one mirror is suspended (released from the substrate for some distance away from anchor points thus permitting specific motions).  In the torsional photonic crystal nanocavity optomechanical sensors designed and operated to date, the moving Bragg mirror is on the end of a short arm extending at right angles from the torsion rod (Fig.~\ref{fig:opticalcavitytransduction}c and d).  The cavity sensitively responds to mirror motion for a set of torsional, flexural, and hybrid torsion/flex modes, the specifics of which depend on all the dimensions and whether the torsion beam is clamped at one or both ends.  The sensitivity can be characterized as a function of wavelength by tuning the probe laser through the cavity resonance.  Such tests reveal both dissipative coupling (quality factor/depth of resonance modulated by mirror position) and dispersive coupling (optical centre frequency of resonance modulated by mirror position).  Optimum combined sensitivity is determined empirically through such tests.  

\begin{figure}[h] 
	\centering
	\includegraphics[width=1\linewidth] {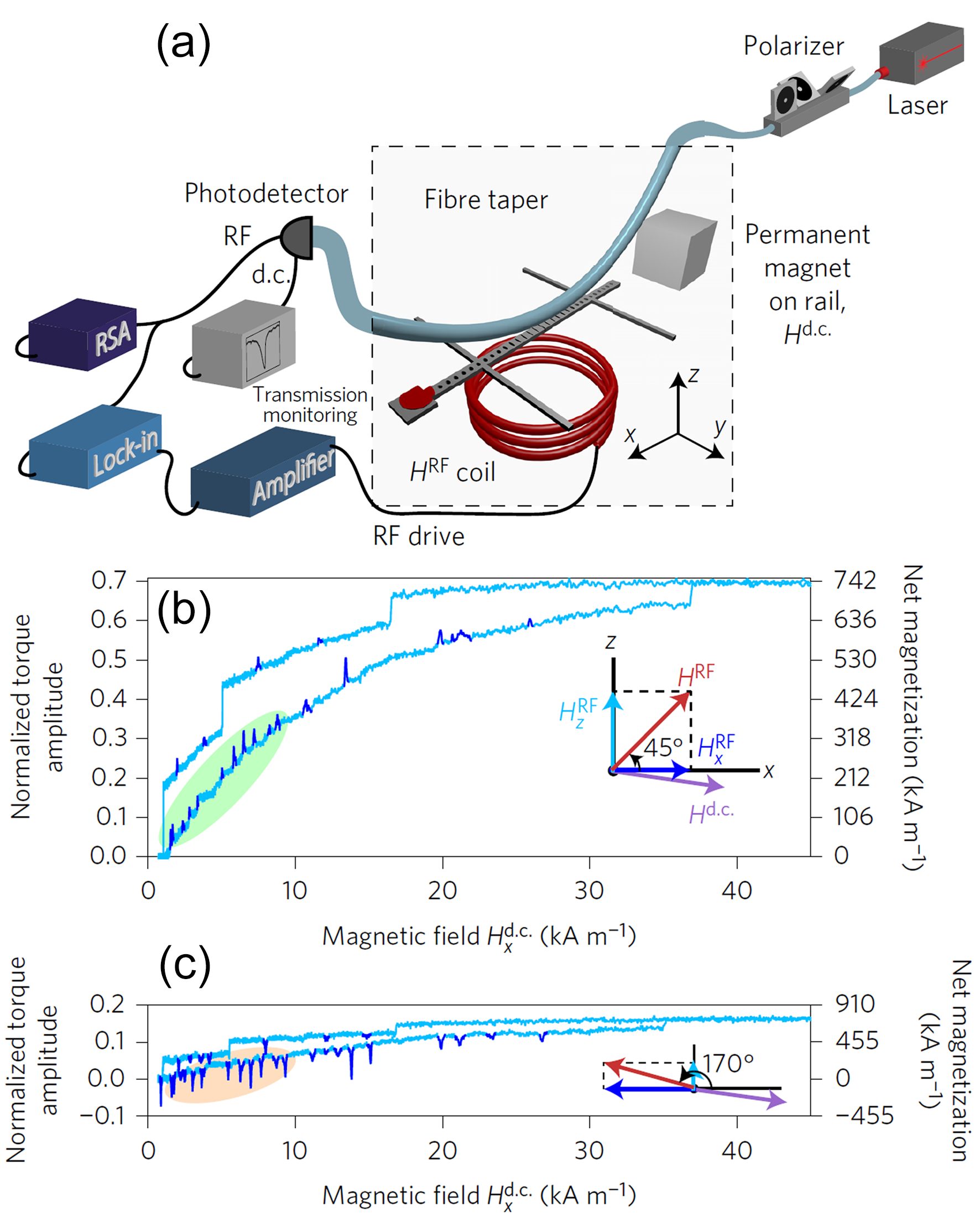}
	\caption{Magnetometry and susceptometry performed with a photonic crystal nanocavity-based nanomechanical torque sensor.  a) A schematic of the measurement setup (not to scale — the device itself is shown $10^4 - 10^5$ times too large, depending on which other component it is compared to).  b,c) A unique investigation enabled by this apparatus was the study of stochastic resonances at Barkhausen steps.  Susceptibility peaks are observed when fast thermally-activated hopping lock-steps with a small in-plane RF magnetic field dither.  The peaks can add to or subtract from the regular magnetometry signal, depending on the relative phases of the in-plane and out-of-plane drive components. (Figures from Ref.[\citenum{Wu2017}], used with permission.)}
	\label{Fig6dot5two}
\end{figure}

In all prototype devices to date, light is coupled in and out evanescently to a tapered optical fibre as illustrated in Fig.~\ref{Fig6dot5two}a. The position of the tapered fibre itself also affects the sensitivity, not only by the strength of optical coupling to the cavity (near critical coupling is most desirable, as discussed in Section \ref{optomech}), but also via the asymmetry and balance of first-order (linear in displacement) and second-order (quadratic in displacement) couplings to mechanical motion\cite{Wu2014}.  It is interesting to note that all of the nanotorsion cavity optomechanical devices operated to date have been transduced through first-order coupling, despite their initial designs having, on paper, very small first-order coupling (only from the asymmetric dielectric environment above and below, shifting the optical mode off-centre).  Further asymmetry is introduced through fabrication procedures (angled sidewalls from silicon etching, small static flexing from strain relieving during oxide etch) and by positioning of the tapered readout fibre with respect to the optical cavity.  Fortunately, the experience has been of the rare and joyful flavour in which things work out even better than anticipated.  

Cavity optomechanical devices have been used to reveal stochastic resonances of vortex cores, driven by weak RF fields between adjacent pinning sites, which are assisted by thermal activation.  These resonances are manifest as small peaks or dips at Barkhausen steps in a hysteresis curve that arises when a small in-plane RF component at the torsion frequency is present, in addition to the out-of-plane RF field dither used for conventional mechanically-resonant torque magnetometry as shown in Fig.~\ref{Fig6dot5two}b, c.  The flexibility of performing measurements in air at room temperature enabled these observations, where the sensitivity loss from air damping of the mechanical resonance was offset by the sensitivity gain of the optical cavity readout.

\section{Torque measurements against the backdrop of ongoing advances in other methodologies}

Experimental prowess continues to advance across the board, and for many studies more than one technique is available.  Different approaches yield related (complementary, but never identical) information.  The sensitivity of torque measurements is competitive with the best other methods across many fronts, with the measurement environment sometimes dictating the optimum selection (for example, a mechanical probe at high magnetic fields or high temperatures where it is difficult to use SQUIDs for magnetometry). The particular applications to which mechanical torque measurements can be declared without hyperbole to be \emph{uniquely} suited (a more stringent criterion than `suitable for') are limited to those defined intrinsically by a torsion mechanical response, such as the determination of magnetomechanical ratios or the observation of spin-lattice relaxation from the perspective of the lattice.  
Direct microscopic probes of spin-mechanical physics are now emerging.  Optical and X-ray studies have been designed and demonstrated to address questions of phonon angular momentum and the internal lattice twisting illustrated in Fig.~\ref{Fig1} and Fig.~\ref{Fig6dot3}.  In a Brillouin light scattering (BLS) experiment, building on decades of beautiful work with BLS to probe magnon transport and other physics \cite{Demokritov2001}, the transport of angular momentum by magnons and phonons over mm distances in YIG was confirmed by optical polarization analysis \cite{Holanda2018} in which angular momentum was pumped into the system by a microwave antenna. This work moves a big step closer to microscopic scales by accessing dispersion curve information about the excitations (including hybridized magnon-phonons).  

Although the BLS measurements are localized within the specimen (in contrast to detecting rigid body response), the effective spatial resolution of the BLS probe remains very large in comparison to atomic scales.  Complementary to this, a time-resolved X-ray diffraction experiment has explored the internal spin-lattice transfer of angular momentum with sub-100 femtosecond resolution.  The experiment effectively initiates the Einstein-de Haas process instantaneously in a thin Fe film, through ultrafast demagnetization with a 40$\thinspace$fs, 800$\thinspace$nm, laser pulse \cite{Beaurepaire1996}.  With the period of spin-lattice relaxation collapsed to an interval of ultrafast brevity in this inspiring stroboscopic study, the transient amplitude of internal twist at the atomic scale (specifically, a transient transverse strain within the film, arising from the impulsive internal torque) becomes large enough to be resolved through X-ray diffraction.  

Other active studies, probing at a more microscopic level the spin-lattice processes that yield conventional magnetic torque sensing as their end result, include work coupling spins magneto-elastically \cite{Jacobsen1959, Pohl1962, Jacobsen1963} to specific phonons introduced as surface acoustic waves \cite{Weiler2011}.  Within the same conceptual vein are studies of magnetic thin films magneto-elastically coupled to highly compliant mechanical structures such as long nanoribbons patterned from LPCVD silicon nitride deposited on thermally grown silicon oxide \cite{Pernpeintner2016}.  The surface acoustic, Rayleigh-type, wave is additionally interesting in this context because of the circular trajectory of particle displacement, which suggests a small Barnett field \cite{Chudnovsky2016, Kobayashi2017}.  Surface acoustic waves in this context become akin to localized sources of microwave magnetic field \cite{Tejada2017, Salasyuk2018}.  
There is also ongoing work towards coupling individual spins in defect centres, such as the nitrogen vacancy in diamond, to local strain.  This work is progressing through strain coupling to orbital states, studied quasistatically \cite{Lee2016} and dynamically \cite{Chen2018}, with strong motivation from the aim of extending the reach to spins and coherent spin-phonon-photon manipulation in hybrid quantum systems \cite{Bassett2014, Kurizki2017}.  

A single experiment has shown indications of more than one aspect of the physics discussed above.  Hashimoto \emph{et al.} \cite{Hashimoto2017} have developed an all-optical method they term spin-wave tomography, which detects broadband magnons and phonons launched in a ferrimagnetic garnet film through absorption of an ultrafast laser pulse.  The excited modes are resolved in dispersion, $(k,\omega)$, and as a function of time delay, through reconstructions of the diffractive Faraday effect imprints on transmitted probe pulses that are polarization-analyzed and imaged with a camera \cite{Hashimoto2017}.  A conclusion from the experiment on lutetium iron garnet is that magnetoelastic coupling dominates the spin-wave generation in small applied magnetic fields, while in large enough fields the effect of optically-induced demagnetization launching spin waves is the most significant.  The detection couples preferentially to magnons, with coupling to phonons highlighted at a dispersion crossing in time-resolved measurements at low fields.

\section{Future projections}

\subsection{Sensitivity limits}

The recent demonstrations of ultrahigh mechanical $Q$’s at room temperature in soft-clamped membrane \cite{Tsaturyan2017} and ribbon resonators \cite{Ghadimi2018} prompts applications to torque sensing.  Such structures could already be applied through torque actuation of flexural modes, and analogous band structure engineering will also be possible for torsion modes.  The Sankey group has already discussed torsional modes in related minimally-supported `trampoline' structures fabricated from silicon nitride membranes \cite{Reinhardt2016, Norte2016}.  The trampolines are studied in optical cavities (`membrane in the middle') and the system stiffness can be dominated by intrinsically dissipation-free optical spring forces (although this is not yet central to the high $Q$’s already demonstrated).  Extraordinary torque sensitivity limits have been theorized for measurements utilizing optical levitation of asymmetric nanoparticles \cite{Hoang2016}, with experimental explorations beginning using double-nanobead dumbbells \cite{Ahn2018}.  In that proposal, one of the most daring to date, the speculated sensitivity limit of $\sim 10^{-29} {\rm N m / \sqrt{Hz}}$ corresponds to the torque on a single Bohr magneton, in a field of 1 A/m (13 mOe), detected in a 1 second observation.  Unlike magnetic resonance force microscopy, where single spin sensitivity was famously demonstrated 1.5 decades ago, a similar potential of pure torque measurements remains to be realized.  The barriers to overcome remain the same as in all past efforts to extend sensitivity: incorporating the specimen of interest without compromising the performance of the system, and suppressing unwanted background signals.  The long-held view of historian Peter Galison from his very in-depth investigation of Einstein-de Haas and Barnett effect experiments (by their namesakes and others) bears quoting in this context, ``Controlling the background is not peripheral to the experimenter’s craft: it is constitutive of the activity itself'' \cite{Galison1987}. It is a blessing to have a sensitive measurement to puzzle through and debug.  

\subsection{Commodity platforms}

The research and development foundry multi-project wafer model \cite{IMEC, A*STAR} has extended the accessibility of custom micro- and nano-optomechanical devices to a wide population of researchers beyond those with their own nanofabrication facilities.  Sophisticated designs, incorporating mechanical sensors coupled to disk and/or ring micro-optical cavies with read-out through integrated optical waveguides, can be sourced in the silicon-on-insulator platform via this route \cite{Chrostowski2015}.  Post-processing is required by the end user, but this can be as little as an oxide etch release \cite{Sauer2012} and/or integration of the magnetic specimen.  (Not to underestimate these tasks, optical lithography to pattern the etch will be necessary to protect the optics when releasing wider mechanical sensors, and a means of depositing or micromanipulating the specimen into position is required.)  State-of-the-art integrated designs incorporating active components such as photodetectors and modulators generally are accessible only through the foundry route, since the complexity exceeds what is feasible within most fabrication facilities in academic settings.  

The silicon photonics foundries use older generation steppers for patterning, equipment forced into retirement by Moore's law driving a requirement for smaller features in integrated circuits.  The dimensional accuracy of small features is not yet at the level needed for photonic crystal designs, but that is coming, driven by the silicon technology roadmap.  Production of state-of-the-art optical nanocavities requires access to 100 keV electron beam writing.  Services have emerged to expand access to this technology as well, but on a more limited scale than the foundries \cite{AppliedNanotools}.

\subsection{Creative opportunities}

The range of viable approaches both to fabrication of torque sensors and to integration of magnetic components now goes well beyond the mostly top-down, subtractive manufacturing strategies emphasized above.  Constraints on what is designed, and for good reasons, are inevitably imposed by what can be machined and modelled.  Both of these constraints are relaxing quickly with the growing combined powers of additive manufacturing and finite element modeling.  Heralding the arrival to torsion mechanics of this new phase of creativity was the recent report of a chiral metamaterial with first-order coupling between compression/extension and twist \cite{Frenzel2017}.   When an isotropic solid material is machined to form a torsion shaft, symmetry precludes any twisting when the shaft is stretched or compressed.  In a spiral spring, the coupling between compression and rotation is second order.  The chiral metamaterial, by contrast, exhibits comparable fractional extensions and rotations.  The opportunites this points to are manifest.  Torque can be transduced in a new way via linear displacement of the spring.  Torsional and flexural properties can be engineered less co-dependently.  The thresholds for nonlinear response to torque may perhaps be engineered to facilitate mechanical parametric amplification.  

Another little-exploited opportunity is represented by the continued emergence of DNA origami, offering possibilities for the design of self-assembling nanoscale mechanical sensors \cite{Zhou2014}.  Independent of the mechanics, self-assembly methods offer possibilities to aid with directed integration of nanomagnetic specimens onto nanoscale sensors \cite{Klein2013}.  The torsion properties of biological molecules are studied as a separate line of inquiry, for their importance to understanding the functioning of biological systems \cite{Lipfert2015}.  

\subsection{Broadband, time-domain methods and applications}

The high displacement sensitivities of cavity optomechanical readouts offer strong prospects for real-time and stroboscopic time-resolved investigations of spin dynamics.  Conventional interferometric readouts for torque magnetometry are able to monitor millisecond scale telegraph noise hopping of vortex cores between adjacent pinning sites \cite{Burgess2013}.  For instance, the greater sensitivity of a $100 \times$ improved cavity readout could result in $10,000 \times$ faster (sub-microsecond) observations with the same resolving power for magnetic moment change.  This increase in measurement bandwidth is obtained easily for (and only for) small signals where the detector response is linear in open-loop operation.  The dynamic range for linear signal transduction remains approximately constant as the cavity $Q$ increases, with the largest signal (at the threshold of optical nonlinearity) scaling down accordingly.  Typical torque signals of interest are very small, and there is a lot of headroom to exploit without introducing fast linearizing feedback schemes.

The rapid ring-up and ring-down times of high $Q$ cavities at optical frequencies (0.2 petahertz at the 1.5 micrometre telecom wavelength standard for silicon-based cavities) also lend themselves to gated optical probing over short intervals \cite{Meenehan2015}.  The concomitant large (even gigahertz) bandwidth enables reasonably fast (to sub-nanosecond resolved) time-domain measurements, with cavity-enhanced sensitivity, to explore transient dynamics and complement continuous wave frequency-domain spectroscopy. In particular, this has application to stroboscopic measurements of repetitive phenomena where boxcar averaging can be used to increase signal-to-noise ratios.  

The mechanical bandwidth of resonant sensors is ordinarily rather low (if not very low), and they are used as slow integrating detectors.  There is a large parameter space open to exploration by combining cavity optomechanical displacement sensitivity with {\em high} mechanical bandwidth sensors (high fundamental torsion frequency with low mechanical $Q$, and/or operated in a passband sufficiently below resonance where the responsivity is approximately frequency-independent).  This approach will have application, for example, to resolving spin-lattice relaxation times from the perspective of the lattice, or in other words by observing the transient buildups and decays of angular motion that are rate-limited by the flow of angular momentum, introduced by magnetic torque applied to the spins, into the mechanical `sink'. 

\subsection{Strongly-coupled cavity spin-mechanical systems}

An enticing prospect is the strong coupling of the magnetization to mechanical degrees of freedom in hybrid sensors.  Proposed by Kovalev \emph{et al.} over a decade ago, magnetomechanical energy coupling grants direct mechanical transduction of magnetic resonances as well as its alluring reciprocal effect, mechanically-assisted magnetization switching \cite{Kovalev2005, Kovalev2006}.  Difficulty in realizing such systems lies in an inadequate overlap of energy scales, where ferromagnetic resonance frequencies (GHz range) don't often coincide with lower MHz frequency nanomechanical modes.  Scaling of mechanical dimensions (often smaller) to allow for higher frequencies often creates additional difficulty in affixing magnetic materials.    

Recent studies employing microwave cavity resonators to strongly coupled systems, through mediation by stored electromagnetic energy, suggests possible methods for magnetomechanical coupling.  In \cite{Tabuchi2014}, magnons in a YIG sphere were strongly coupled to superconducting qubits, allowing for their coherent parametric control.  A similar magnonic system was also coherently coupled to photons and phonons \cite{Zhang2016}. 

\section{Summary}

The increasing capabilities to control and couple excitations in materials, exemplified by strong phonon-photon interactions in nano-optomechanical cavities, are transformative in many domains including mechanically-based studies of magnetism.  Part of a larger blossoming in the design and fabrication of metamaterials with special physical properties is introducing unanticipated, significant, and quantitative leaps in the ability to reduce dissipation in small mechanical resonators.  What might be considered routine torque sensitivity from mechanical sensors has improved since the time of John Moreland's review \cite{Moreland2003} from $\sim 10^{-17} {\thinspace\rm N  m/\sqrt{Hz}}$ to $\sim 10^{-20} {\thinspace\rm N  m/\sqrt{Hz}}$, which would surpass his projected limit of MEMS magnetic moment sensitivity by perhaps an order of magnitude. These advances will continue, with record reported torque sensitivities already more than three orders of magnitude below the `routine' benchmark.  

For reference to spin sensitivity, it is helpful to reflect on the fundamental mechanisms underlying different detection methods.  In conventional inductive detection of spin resonance, the improved sensitivity at higher fields and frequencies is the result of higher RF photon energies (as well as increased statistical polarization, in the case of paramagnetic resonances).  In contrast, for pure torque detection of spin resonance through Einstein-de Haas mechanisms, the increment of angular momentum transferred per transition is fixed.  A torque sensitivity of $10^{-24} {\thinspace\rm N  m/\sqrt{Hz}}$ is sufficient to detect $10^{10}$ spin-$\frac{1}{2}$ transitions per second (which can be parsed in any way between the number of spins and the individual transition rates, for example $10^5$ spins at $10^5$ flips per second).  

Owing to the large range of materials and relaxation rates of interest, there is great interest in expanding the range of mechanical sensor resonance frequencies and, independently, damping rates, which are currently in the MHz range for $f_{\rm mech}$ and the sub-kHz range for the linewidth, $\Delta f_{\rm mech}$, to a range from $\lt 1\thinspace$Hz to $\gt 10^9 \thinspace$Hz for both $f_{\rm mech}$ and $\Delta f_{\rm mech}$ (again, independently).  Higher phonon energies are one reason why higher mechanical frequency structures are more difficult to transduce, but a compensating factor for optomechanics vs. inductive measurements is that GHz phonons in \newpage
\noindent a resonator are much more localized than GHz photons, owing to low acoustic velocity (and comparable instead to 100 THz photons).  There has even been remarkable progress towards inductive detection of single electron spin resonance, a goal that wasn't on the horizon 15 years ago \cite{Probst2017}.  If there is one take-away to summarize progress, it is that people are relentlessly ingenious.

\section*{Acknowledgements}
The authors wish to acknowledge support from the Natural Sciences and Engineering Research Council, Canada, the Canada Research Chairs program, and the Faculty of Science, University of Alberta.  The research described in Sections \ref{optomech}, \ref{magnetic_hysteresis}, and \ref{spin_resonance} was supported additionally by the National Institute for Nanotechnology, National Research Council, Canada.  We thank Michael Dunsmore, Ghazal Haji salem, Danielle Jenson, Gustavo O. Luiz, and John Thibault for careful readings of the manuscript and their feedback as well as David Fortin for the rendering of Fig.~\ref{Fig1} and Kayte Mori for assistance in data acquisition for Fig.~\ref{Fig6dot2}.

\newpage

\clearpage

\section*{Appendix: Designer's guide}

\renewcommand{\theequation}{A\arabic{equation}}
\setcounter{equation}{0}

A defining characteristic of mechanical torque-based studies of magnetic phenomena is the large scalability of the sensors and their sensitivities to suit the investigation of interest.  LIGO \cite{Abbott2016} has set such a high bar for fractional displacement sensitivity and extreme noise reduction (including seismic isolation to a part in a trillion of ground vibration) that the small-scale laboratory experimentalist should be undeterred by, and mindful of, any analogous challenges.  

\subsection*{A feel for the numbers}

To establish the context for discussion, the scales of mechanical torques arising from different magnetic mechanisms are compared to torque sensitivities readily achievable with present nanoscale torsion sensors operating at room temperature.  We choose $1 \times 10^{-22}\thinspace {\rm N  m /\sqrt{Hz}}$ as the sensitivity benchmark, representing the next iteration beyond the vacuum sensitivities of initial micro- and nanocavity optomechanical devices as described in \citenum{Kim2017, Wu2017} and discussed further in Section 6.5.  In a standard magnetometry, this corresponds to the torque on a moment equivalent to $10^5$ Bohr magnetons in an applied field of 0.1$\thinspace$kA/m.  For a study of magnetic anisotropy, this is the scale of peak AC torque from magnetocrystalline anisotropy on a 50$\thinspace$nm diameter YIG sphere biased in a DC field of 100$\thinspace$kA/m and dithered at right angles by RF fields with amplitude 0.1$\thinspace$kA/m.  In angular momentum pumping experiments, this corresponds to a $dL/dt$ torque of $10^{12}$ spin-$\frac{1}{2}$ flips per second, or about 150 nA of spin polarized current into a spin sink.  For photon pumping, an excitation level of $10^{12}$ transitions per second corresponds to a power absorption (or circularly polarized emission) of 6.7$\thinspace$pW for 10$\thinspace$GHz microwave photons (3$\thinspace$cm wavelength, 6.7$\thinspace$yJ or 42$\thinspace \mu$eV energy), a power absorption (or emission) of 240$\thinspace$nW for 830$\thinspace$nm near-infrared photons (360$\thinspace$THz, 240$\thinspace$zJ or 1.5$\thinspace$eV energy), or comparably a 6.7$\thinspace$pW flux of 10$\thinspace$GHz magnons.  

\subsection*{Underlying scaling behaviour as a function of torque sensor/specimen linear size}
\renewcommand{\thefigure}{A\arabic{figure}}
\setcounter{figure}{0}

\begin{figure}[ht] 
	\centering
	\includegraphics[width=1\linewidth] {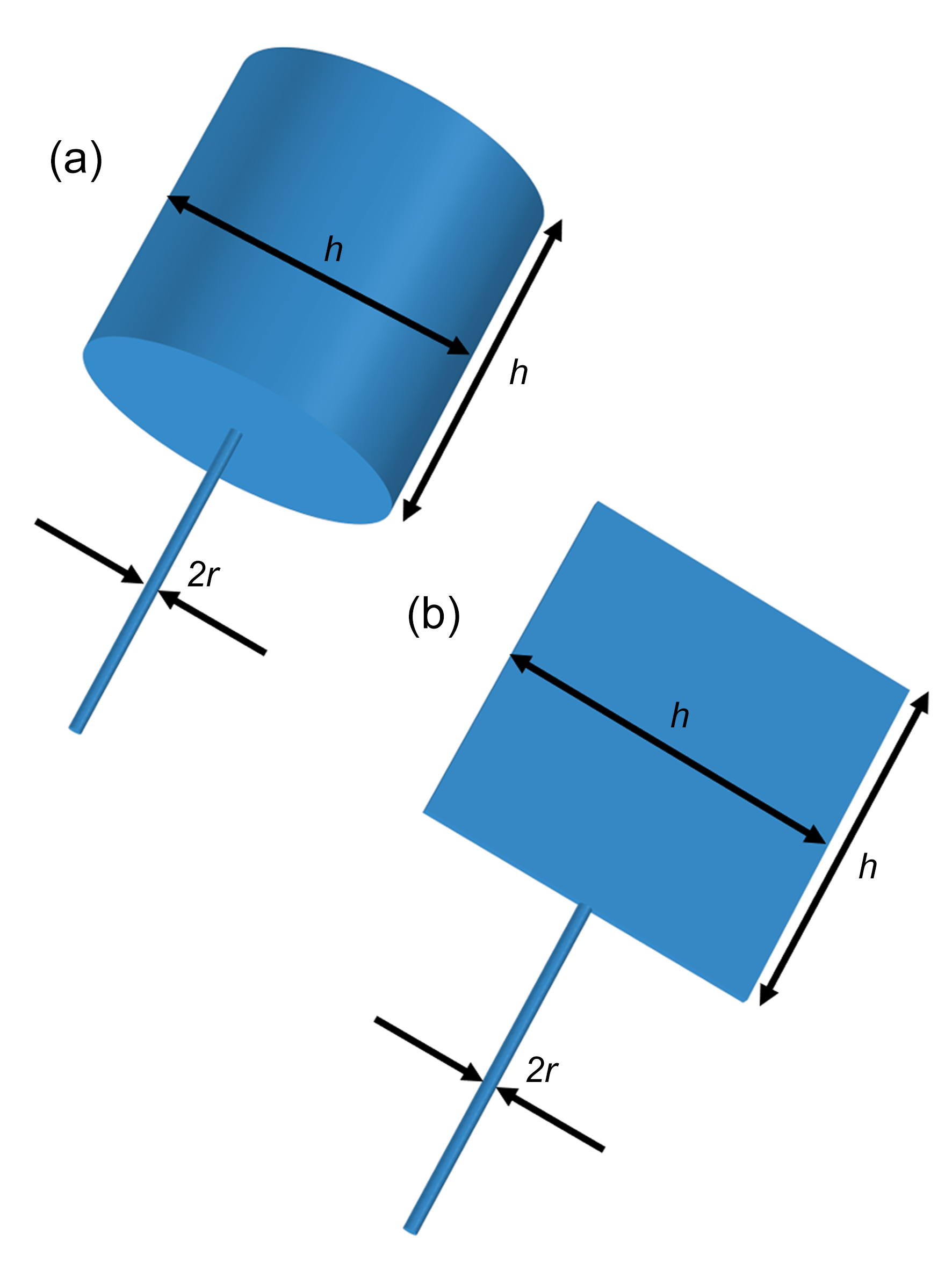}
	\caption{Simple torsional resonator basis structures. Cylinder (a) and paddle (b) affixed to torsion rods.}
	\label{FigA}
\end{figure}
Fig.~\ref{FigA} presents two generically shaped torsional resonators, one with a cylinder and the other having a paddle as the pendulum bob.  We define the dimensions through an aspect ratio $A = h/r$, so that the scaling behaviour is a function only of torsion rod radius, $r$.  The torsion constants, $\kappa$, of both devices scale as $r^4/h = r^3/A$ (the counter-torque from the development of shear stress in the rod, arising in response to twist, decreases with radius as $r^4$, but the spring also becomes shorter and hence stiffens as $1/r$).  The moments of inertia, $I$, of the pendulum bobs, assumed to include the magnetic specimen of interest and to have uniform densities, scale as $I \propto A^2r^5 \rho_{\rm volume}$ for the cylinder and $I \propto A^2 r^4 \rho_{\rm area}$ for the paddle.  The fundamental torsional resonance frequencies of the two devices therefore scale as $f_0 \propto 1/r$ for the cylinder, and $f_0 \propto 1/\sqrt{r}$ for the paddle, with an additional multiplicative factor $A^{-\frac{3}{2}} \rho^{-\frac{1}{2}}$ common to both cases (with $\rho = \rho_{\rm volume}$ or $\rho_{\rm area}$ as appropriate).  

Inserting these results into the expression for thermomechanically-limited torque sensitivity, 
\begin{equation} \label{}
\tau_{\mathrm{min}}^{\mathrm{(1 Hz\thinspace BW)}} = \sqrt{\frac{4 k_B T}{\pi}\frac{\kappa_{\mathrm{eff}}}{Q f_0}}
\end{equation}
we find that (taking $T$ and $Q$ to be constants):
\begin{equation} \label{}
\tau_{\rm min}^{\rm cylinder} \propto r^2(A \rho_{\rm volume})^{1/4}, {\rm \: and \:}
\tau_{\rm min}^{\rm paddle} \propto r^{7/4}(A \rho_{\rm area})^{1/4}.
\end{equation}
The take-away from these expressions is that for the most frequent scenario of 3D specimens where the torque signal scales with sample volume and hence is presumed to vary with $r^3$, the simple constant aspect ratio geometric scaling compensates for only two of the powers of $r$ and highly detailed engineering in sensor design and fabrication is required to compensate for the third power of $r$ in order to maintain constant signal-to-noise ratio (SNR) in a 1 Hz bandwidth, as sizes decrease.  A commonly used partial work-around is to increase measurement times.  Conversely, with all else being equal, SNR will increase as linear dimensions (including sample volume) are increased.  Overall, there is some allowance for parameter variation in sensor design for comparable performance provided the sensor and specimen are approximately matched.  Note that this very basic scaling analysis predicts a weak dependence on aspect ratio, with sensitivity varying slowly as $A^{1/4}$.  In practice, $Q$ will also depend on $A$, decreasing at small $A$ and somewhat offsetting this weak dependence.  The most significant take-away regarding the dependence on $A$ is that for specimens roughly filling the cylinder or paddle, sample volume also scales as $A^3$ or $A^2$; this is of particular significance to designs for weakly-magnetic samples.  Yet another advantage of larger $A$ is that the same angular displacement translates to a increased linear displacement at the edge of the paddle, making it easier to reach thermomechanically-limited displacement detection with the interferometric readout. 

\subsection*{Readout sensitivity}

We turn now to the laser and photodetector requirements for a sensitive interferometric readout, with displacement noise chosen somewhat arbitrarily to be $10 \times$ below the Brownian thermomechanical amplitude.  It is helpful to be able to detect the resonance modes readily through Brownian motion alone as measured with a spectrum analyzer, without any driving torque.  The limiting amplitude of Brownian motion at frequencies below resonance is reduced by $1/Q$.  Section 8.4 touches upon applications, such as high mechanical bandwidth measurements, that can benefit from being thermomechanically-limited off-resonance.

In general, we would like to know the lowest optical power with which provides a required desired displacement resolution and which minimizes heating of the sensor caused by the measurement.  The thermomechanical angular amplitude spectral density at resonance is 
\begin{equation} {}
S_{\theta}(f_0) = \sqrt{\frac{2k_B T Q}{\pi\kappa f_0}}\thinspace {\rm rad/\sqrt{Hz}}
\end{equation}
Substituting the scaling relations for the paddle shape (torsion rod radius $r$, aspect ratio $A$), we find that the thermomechanical displacement amplitude spectral density at the end of the paddle,
\begin{equation} {}
S_{z}(f_0) = \sqrt{\frac{2k_B T Q}{\pi\kappa f_0}} \frac{Ar}{2}\thinspace {\rm m/\sqrt{Hz}},
\end{equation}
scales as 
\begin{equation} {}
S_{z}(r,A) \propto A^{9/4} \rho_{\rm area}^{1/4} r^{-1/4}.
\end{equation}

The peak Brownian motion displacement amplitude, $S_{z}$, in the fundamental torsional mode (at $f_0$ and measured at the edge of the paddle) scales very slowly with overall size, increasing as $r^{-1/4}$ as $r$ is reduced.  (In principle, $S_z$ is $r$-independent for the cylindrical resonator, all else being equal.)  $S_z$ varies rapidly with aspect ratio $A$, as $A^{9/4}$, with a contribution of one power of $A$ from the length of the mechanical lever arm to the edge of the paddle, $A^{3/4}$ from $f_0$ decreasing with $A$, and $A^{1/2}$ from $\kappa$ decreasing as the torsion spring lengthens.  

In practice, at room temperature, for typical aspect ratios of 10-20 and modest mechanical $Q$s of order 1000, we find that on resonance $S_z \sim 0.5\thinspace {\rm pm/\sqrt{Hz}}$.  As a specific example, this corresponds to a target displacement sensitivity $z_{\rm meas}^{\rm (1 Hz BW)} = 0.05\thinspace {\rm pm/\sqrt{Hz}}$.

\subsubsection*{Low finesse interferometer}

For interferometric displacement detection measured in reflection, and with a given incident optical intensity, $I_0$, the reflected signal from a Fabry-Perot interferometer is $I=I_0 R_{\rm FP}$, where $R_{\rm FP}$ is the reflectivity of the interferometer.  Figure A2 illustrates the variation in reflectance, $R_{\rm FP}$, as a function of mirror spacing (dotted lines), and corresponding relative sensitivity to small displacements ($dR_{\rm FP}/dz$, solid lines) for two `non-cavity' Fabry-Perot interferometers typical of nanomechanical displacement detection; or in other words, interferometers having finesse, $\mathcal{F}$, not large compared to 1 on account of low component mirror reflectivities, $R_{\rm m} \ll 1$ (assumed equal for both mirrors).    
\begin{figure}[h] 
	\centering
	\includegraphics[width=1\linewidth] {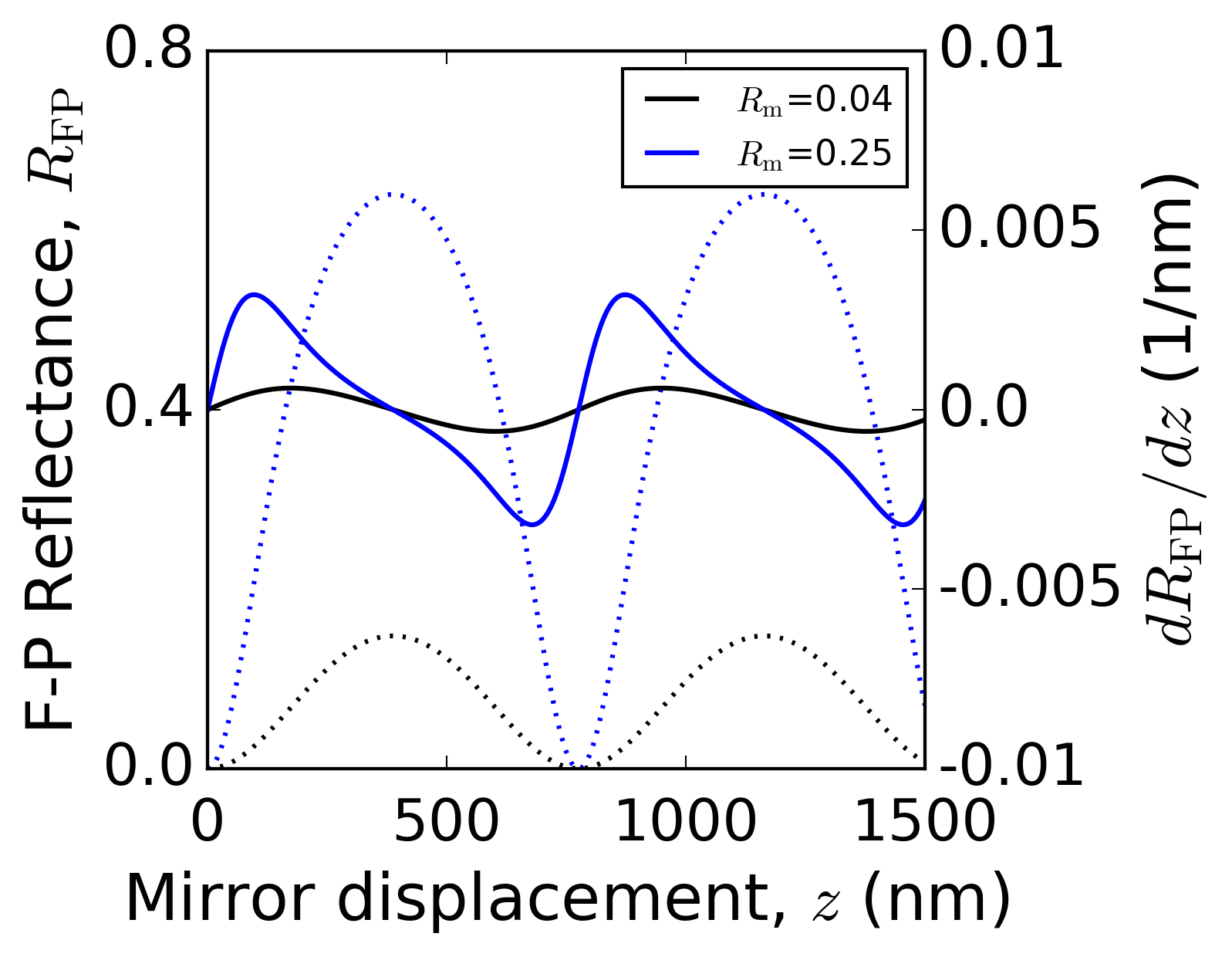}
	\caption{Reflectance of a Fabry-Perot cavity versus displacement of the moving mirror, for incident 1550$\thinspace$nm radiation and for two different mirror reflectances, $R_{\rm m} \ll 1$.  The dotted lines show the cavity reflectance versus displacement.  The solid lines are the first derivative, $dR_{\rm FP}/dz$ (scale on the right axis), and capture the relevant performance of the interferometer as a detector of small displacements.  The two mirror reflectances shown correspond to cavity finesses $\mathcal{F} = 0.65 (R_{\rm m} = 0.04)$ and $\mathcal{F} = 2.09 (R_{\rm m} = 0.25)$.}
	\label{FigA2}
\end{figure}

$R_{\rm m} \ll 1$ corresponds to an interferometer where the input mirror is an uncoated optical fibre cleave and the distance to the back reflector has been adjusted to optimize the modulation depth and yield interference fringe visibility $V = (R_{\rm FP}^{\rm max} - R_{\rm FP}^{\rm min}) / (R_{\rm FP}^{\rm max} + R_{\rm FP}^{\rm min}) \approx 1$.  This adjustment makes the fraction of power collected from the back mirror approximately $0.04$, through a combination of actual higher back mirror reflectance and power loss from diffractive spreading of the beam.  Higher laser power is often employed for increased sensitivity in this case, unless the unused, scattered radiation causes unwanted heating or other problems.  $R_{\rm m} = 0.25$ is typical of a silicon-on-insulator, free-space coupled device.

At $R_{\rm m} \ll 1$, the maximum slope of interferometer reflectance vs. $z$ is 
\begin{equation} {}
\left|\frac{dR_{\rm FP}}{dz}\right|_{\rm max} = \frac{8\pi}{\lambda}R_{\rm m},
\end{equation}
where $\lambda$ is the probe wavelength.  For small displacements we are interested in the fractional change of intensity, $\Delta I / I_{\rm DC}$ (where $I_{\rm DC}$ is the DC reflected optical intensity at this operating point), which must be resolved in order to detect a displacement $z_{\rm meas}$.  The fractional change will let us determine the minimum detected power to avoid being dominated by optical shot noise (root mean square statistical fluctuations in photon number of $\sqrt{2N}$ per second for a power corresponding to $N$ photons per second \cite{Edelstein1978}), and photodetector noise, combined in quadrature.  At visibility $V=1$, the reflectance modulates with interference between 0 and $4R_{\rm m}$, such that $I_{\rm DC} = I_0(2R_{m})$.  Hence,
\begin{equation} {}
\frac{\Delta I}{I_{\rm DC}} = \frac{4\pi}{\lambda} \Delta z
\end{equation}
Taking $\Delta z = z_{\rm meas} = 0.05\thinspace{\rm pm}\thinspace$ and $\lambda = 1.5\thinspace \mu$m (the wavelength range where narrow-line tunable diode lasers are comparatively inexpensive thanks to telecommunications technology), we find $\Delta I/(I_{\rm DC}) = 0.4 \times 10^{-6}$.  To be shot noise limited in a 1 second bandwidth at this resolution we require $2(1/0.4 \times 10^{-6})^2 = 1.3 \times 10^{-13}$ photons/s, which at 130$\thinspace$zJ/photon is 1.6$\thinspace \mu$W.  The optical shot noise power is $\sim 0.44\thinspace {\rm pW/\sqrt{Hz}}$, approximately the same as the noise equivalent power of the photo-receivers often used that have dark noise (Noise Equivalent Power, NEP referred to the input) of $\sim 0.4\thinspace {\rm pW/\sqrt{Hz}}$ \cite{NewFocus2053}.  A minimum laser intensity yielding $I_{\rm DC} \sim 2.2 \thinspace \mu$W at the detector should provide the desired SNR above the combined detector and shot noise, when $R_{\rm m} \ll 1$.  The circumstances of Fig. A2 are analogous to the integrated Mach-Zehnder interferometer of Fig. 5.     

\subsubsection*{High finesse optical cavity}
\begin{figure}[h] 
	\centering
	\includegraphics[width=1\linewidth] {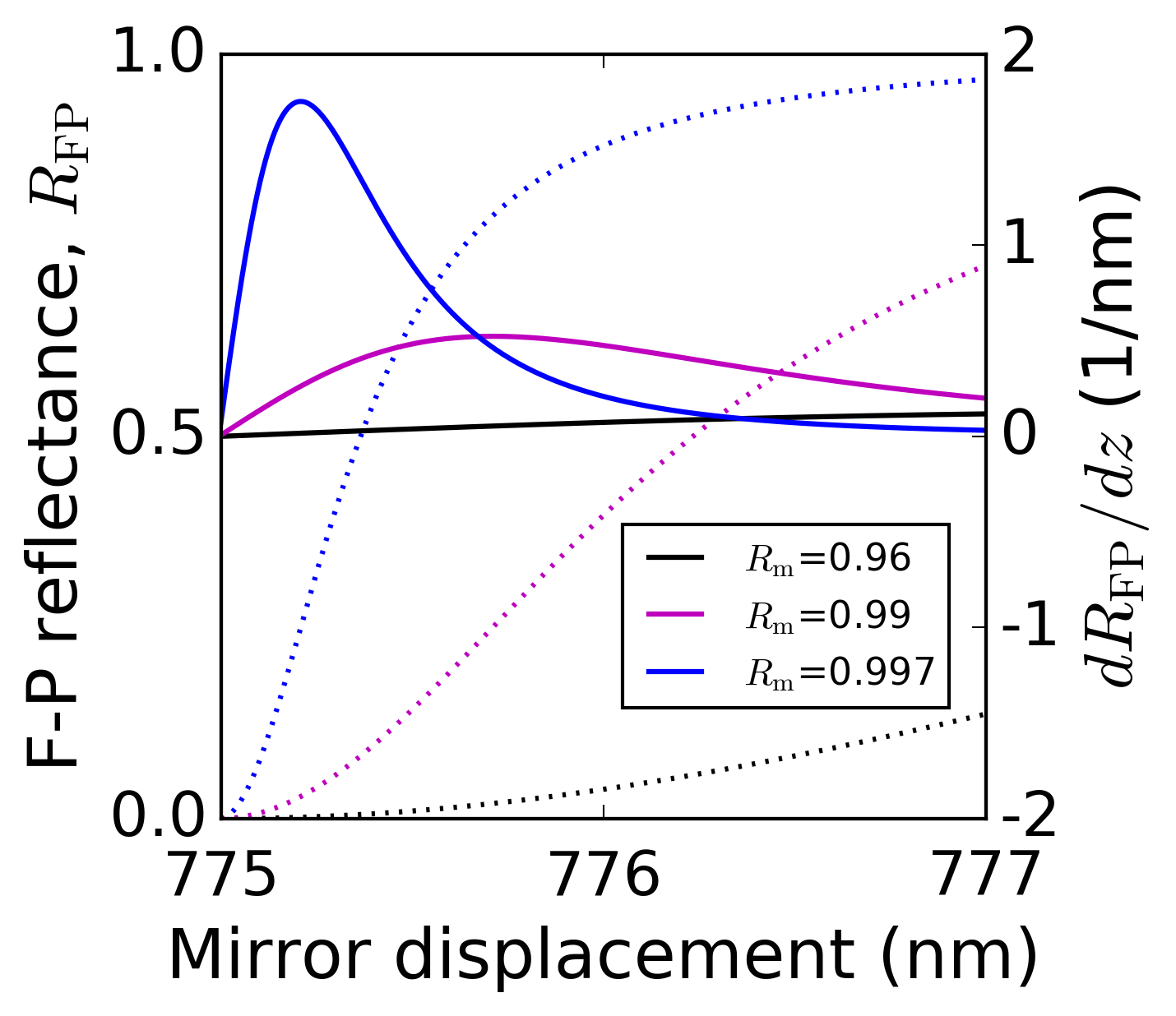}
	\caption{Zooming in on the sensitive detection region for higher finesse cavities, again for 1550$\thinspace$nm wavelength probe light.  Three mirror reflectances are shown, corresponding to cavity finesses $\mathcal{F} = 76.9$  ($R_{\rm m}$ = 0.96),  $\mathcal{F} = 312$  ($R_{\rm m}$ = 0.99), $\mathcal{F} = 1045$  ($R_{\rm m}$ = 0.997).  Note that the scale of the $x$-axis here is $750 \times$ less and of the $y-$right axis $200 \times$ more than in Fig. A2.}
	\label{FigA3}
\end{figure}
Figure A3 illustrates the sensitivity enhancements when cavity finesse, $\mathcal{F} \gg 1$.  The maximum obtainable sensitivity increase scales as  $2\mathcal{F}/\pi$\cite{Shoemaker1991}, and 
\begin{equation} {}
\frac{\Delta I}{I_{\rm DC}} = \frac{8\mathcal{F}}{\lambda} \Delta z.
\end{equation}
Taking $\Delta I =$ NEP again, we find the minimum detection power $I_{\rm DC} \sim 2\thinspace \mu{\rm W} \times \pi/2\mathcal{F}$ (with corresponding optical shot noise smaller than at low $\mathcal{F}$).  The increased sensitivity arises on account of each probe photon, on average, making $\sim \mathcal{F}$ round trips inside the optical cavity, thus accumulating phase shifts.  Note that as the minimum required detected power level $I_{\rm DC}$ decreases as $1/\mathcal{F}$, the power inside the cavity itself is unchanged.  A minimum number of photon encounters with the movable mirror per unit time are required in this measurement to extract the desired information\footnote{It is possible that different scaling behaviour will emerge for measurements with correlated photon pairs; see for example Ref.[\citenum{Lyons2016}].}.  The mirror reflectivities chosen for Fig. A3 are characteristic of metallic ($R_{\rm m} = 0.96$) and broadband dielectric mirrors ($R_{\rm m} = 0.99\thinspace {\rm and}\thinspace 0.997$).  These cases are to be compared with the integrated racetrack resonator of Fig. 5.  

A rule of thumb for the maximum peak-to-peak displacement that can be measured with linear optomechanical response for a given mirror reflectance is approximately $50(1-R_m)\thinspace$ nm or $5 \times 10^4 (1-R_m)\thinspace$ pm (somewhat larger for small $R_m$).  In practice, the laser wavelength is tuned to a sensitive operating point. If $\mathcal{F}$ is increased enough, a point is reached where frequency fluctuations of the laser become a limiting source of noise, and laser stabilization is needed.  The tuning resolution and laser frequency stability required increase with the length of the Fabry-Perot cavity.  Just as significantly, if one chooses to store more optical energy in the cavity by increasing $\mathcal{F}$, a crossover can be reached beyond which radiation pressure fluctuations induce larger random motion of the resonator position than caused by the thermal energy. 



\bibliographystyle{unsrtnat}
\bibliography{refs-v10-06_14_2018}




  
\hfill\break
\end{document}